\documentclass[apj]{emulateapj}
\usepackage{comment} 
\usepackage{enumerate}

\shorttitle{ICM-ISM EFFECTS ON THE FIR-RADIO CORRELATION}
\shortauthors{MURPHY ET AL.}

\slugcomment{Draft version 2.2 \today}
\journalinfo{Submitted to \apj, June 9, 2008}

\begin{document}

\title{Environmental Effects in Clusters: Modified Far-Infrared--Radio
  Relations within Virgo Cluster Galaxies} 

\author{
E.J.~Murphy,\altaffilmark{1,2} J.D.P.~Kenney,\altaffilmark{1} 
G.~Helou,\altaffilmark{3} A.~Chung,\altaffilmark{4} and J.H.~Howell\altaffilmark{2}}
\altaffiltext{1}{\scriptsize Department of Astronomy, Yale University,
  P.O. Box 208101, New Haven, CT 06520-8101}
\altaffiltext{2}{\scriptsize {\it Spitzer} Science Center, MC 314-6,
  California Institute of Technology, Pasadena, CA 91125; emurphy@ipac.caltech.edu}
\altaffiltext{3}{\scriptsize California Institute of Technology, MC
  314-6, Pasadena, CA 91125}
\altaffiltext{4}{\scriptsize Jansky Fellow of the NRAO at the University of Massachusetts, Amherst, MA 01003}

\begin{abstract}
We present a study on the effects of the intracluster medium (ICM) on
the interstellar medium (ISM) of 10 Virgo cluster galaxies 
using {\it Spitzer} far-infrared (FIR) and VLA radio continuum imaging.
Relying on the FIR-radio correlation within normal galaxies, we use our
infrared data to create model radio maps which we compare to the
observed radio images. 
For 6 of our sample galaxies we find regions along their outer edges
that are highly deficient in the radio compared with our models.
We also detect FIR emission slightly beyond the observed radio disk along 
these outer edges. 
We believe these observations are the signatures of ICM ram pressure.  
For NGC~4522 we find the radio deficit region to lie just exterior to a region of high radio polarization 
and flat radio spectral index, although the total 20~cm radio continuum in this region does not appear strongly enhanced. 
These characteristics seem consistent for other galaxies with radio polarization data in the literature.  
The strength of the radio deficit is inversely correlated with the time since peak pressure as inferred from stellar population studies and gas stripping simulations, suggesting the strength of the radio deficit is good indicator of the strength of the current ram pressure.   
We also find that galaxies having {\it local} radio {\it deficits} appear to have {\it enhanced global} radio fluxes.  
Our preferred physical picture is that the observed radio deficit
regions arise from the ICM wind sweeping away cosmic-ray (CR) electrons
and the associated magnetic field, 
thereby creating synchrotron tails as observed for some of our galaxies. 
We propose that CR particles are also re-accelerated by ICM-driven shocklets behind the observed
radio deficit regions which in turn enhances the remaining radio disk brightness.  
The high radio polarization and lack of precisely coincident enhancement in the total synchrotron power for these regions suggests shearing, and possibly mild compression of the magnetic field, as the ICM wind drags and stretches the leading edge of the ISM.   
\end{abstract}
\keywords{galaxies: clusters: individual (Virgo) --- infrared: galaxies --- radio continuum: galaxies --- cosmic-rays --- galaxies: interactions --- galaxies: ISM} 

\section{Introduction}
The physical processes associated with interactions between the
intracluster medium (ICM) and the interstellar medium (ISM) play
critical roles driving the evolution of spiral galaxies in clusters
\citep[e.g.][]{gg72,ltc80,amb99,ss01,bv01}.  
Many basic effects related to ICM-ISM interactions (e.g., ram pressure stripping) are still not well understood.  
These effects include the fate of star-forming molecular clouds, the
rates of triggered star formation  versus gas removal, and the
possible reconfiguration 
of a galaxy's large-scale magnetic field and/or cosmic-ray (CR) particles.  
It is also not known how far from cluster centers the stripping
occurs; 
for example, gas-deficient galaxies are found in cluster outskirts
\citep{sol01} yet it remains unclear if they are stripped in cluster
cores or locally in the outskirts as seems to be the case in NGC~4522 \citep{kvv04}.  

If we can find clear tracers of ongoing ram pressure and identify galaxies in different stages of the stripping process, then we can further understand the consequences of stripping.
Since ram pressure will more easily affect lower density constituents
of the ISM, it seems that the diffuse radio continuum halos of
galaxies may be sensitive tracers of active ICM pressure as indicated
by observations for a number of Virgo cluster spirals including 
NGC~4522 \citep{bv04}, NGC~4402 \citep{hc05}, NGC~4254 \citep{kc07}, 
and a few others \citep{bv07}.  
More specifically, large regions of enhanced polarized radio continuum
emission have been found within a number of cluster spirals
\citep{bv07, kc07}; 
the maxima of the polarized radio continuum within these
galaxies are located along outer edges and thought to arise from external influences of the cluster environment.  
However, without a good idea of the galaxy's unperturbed appearance in
the radio, these observations alone make it difficult to quantify the
ram pressure effects.  
We now address this by predicting the radio distribution expected in the absence of ram pressure by
using the nearly universal correlation between the far-infrared (FIR) and non-thermal radio continuum emission of normal galaxies \citep[e.g.][]{de85,gxh85}. 

Studies of the FIR-radio correlation within field galaxies \citep[e.g.][]{hoer98,hip03,ejm06a} have found that the dispersion in the FIR/radio ratios across a galaxy disk is as small as the dispersion of the global correlation (i.e. $\la$0.3~dex) on scales  down to $\sim$100~pc.  
While a number of other authors have studied how global FIR/radio ratios behave among cluster galaxies \citep[e.g.][]{mo01,ry04}, finding them to be systematically lower than those for field galaxies, it is only now by using {\it Spitzer} that we can study the influence of the cluster environment on the FIR/radio behavior {\it within} galaxies allowing us to better explain the global FIR/radio observations.

The FIR-radio correlation is driven by the process of massive star formation.
Massive stars both heat dust, yielding FIR emission, and end their lives as supernovae (SNe) whose remnants (SNRs), via shocks, accelerate CR electrons in a galaxy's magnetic field bringing about synchrotron radiation.
Using a phenomenological smearing model, in which a galaxy's FIR map is
smoothed by a parameterized kernel to compensate for the fact that the
mean free path of dust heating photons is much shorter than the
diffusion length of CR electrons, \citet{ejm08} has shown for a sample
of 15 non-Virgo spiral galaxies that the dispersion in the FIR/radio ratios
on sub-kiloparsec scales {\it within} galaxies can be reduced by a
factor of $\sim$2, on average.    
Accordingly, it is possible to derive a good first order approximation
of an undisturbed galaxy's non-thermal radio continuum morphology from
its FIR image. 

Assuming that dust and gas are well mixed within galaxies, 
the FIR emission distribution provides a reasonable picture for the molecular gas distribution as modulated by heating by starlight.  
The non-thermal radio continuum emission, on the other hand, is a function of a galaxy's magnetic field and CR electron distributions (i.e. the relativistic phase of the ISM).  
Systematic departures from the nominal FIR/radio ratios across galaxy disks undergoing ICM-ISM interactions can provide insight into differential influence of these interactions on the gaseous and relativistic phases of the ISM.  
These differences may help to quantify the ICM pressure acting on the ISM of such galaxies.  

With the capabilities of current observatories the Virgo cluster
remains an ideal laboratory allowing for many resolution elements
across a large number of galaxy disks, especially at FIR wavelengths.
Coupling FIR observations taken by the {\it Spitzer} Space Telescope
with VLA radio continuum imaging, obtained as part of the VLA Imaging
of Virgo in Atomic gas (VIVA; A. Chung et al. 2008, in preparation)
survey, we study how the relativistic and gaseous phases of the ISM are affected by ICM-ISM 
interactions for a sample of Virgo cluster galaxies. 
This is done using FIR {\it Spitzer} maps to predict how the
radio morphology should appear if the galaxy were not experiencing an interaction; 
in this paper we make the case that significant deviations from such
an appearance are directly related to ICM-ISM interactions.

The paper is organized as follows: 
In $\S2$ we briefly describe the galaxy sample and observations.
We then present our analysis techniques for comparing the FIR and
radio continuum imaging in $\S$3.
Results from this comparison are presented in $\S$4.
In $\S$5 we compare our results in detail with other observational
evidence of ICM effects for NGC~4522, a well studied galaxy
experiencing ram-pressure stripping, and put together a physical
scenario for our observations.  
Then, in $\S$6, we compare and discuss this physical picture for the
rest of the sample.  
Finally, we summarize our conclusions in $\S$7.

\begin{deluxetable*}{ccccccccccccc}
  \tablecaption{Basic Galaxy Data\label{tbl-galdat}}
  \tablehead{
    \colhead{} & \colhead{R.A.} & \colhead{Decl.} &
    \colhead{} & \colhead{} & \colhead{} & 
    \colhead{$M_{B}$} & \colhead{$V_{r}$} & 
    \colhead{$D_{25}$} & \colhead{} & \colhead{PA} & 
    \colhead{$d_{\rm M~87}$} & \colhead{H~{\sc i}~Def.}\\
    \colhead{Galaxy} & \colhead{(J2000)} & \colhead{(J2000)} &
    \colhead{Type} & \colhead{H$\alpha$~Type} & \colhead{Nuc.} &
    \colhead{mag} & \colhead{(km s$^{-1}$)} & 
    \colhead{(arcmin)} & \colhead{$b/a$} & \colhead{($\degr$)} &
     \colhead{($\degr$)} & \colhead{(dex)} \\ 
    \colhead{(1)} & \colhead{(2)} & \colhead{(3)} & \colhead{(4)} & 
    \colhead{(5)} & \colhead{(6)} & \colhead{(7)} & \colhead{(8)} &
    \colhead{(9)} & \colhead{(10)} & \colhead{(11)} & \colhead{(12)} &
    \colhead{(13)}}
  \startdata
NGC~4254&12 18 49.4 &+14 25 07 & SAc	  &N	   & \nodata&-20.66 & 2407& 5.4& 0.87&   0   & 3.6&-0.10\\
NGC~4321&12 22 55.2 &+15 49 23 & SABbc    &T/N     & L/H    &-21.05 & 1571& 7.4& 0.85&  30   & 4.0& 0.35\\
NGC~4330&12 23 16.5 &+11 22 06 & Scd	&T/N $^{a}$& \nodata&-18.01 & 1565& 4.5& 0.20&  59   & 2.1& 0.80\\
NGC~4388&12 25 47.0 &+12 39 42 & SAb &T/N~[s]$^{a}$& Sy2    &-19.34 & 2524& 5.6& 0.23&  92   & 1.3& 1.16\\
NGC~4396&12 25 59.3 &+15 40 19 & SAd	  &N$^{a}$& \nodata&-18.04 & -128& 3.3& 0.29& 125   & 3.5& 0.30\\
NGC~4402&12 26 07.9 &+13 06 46 & Sb	 &T/N$^{a}$& \nodata&-18.55 &  232& 3.9& 0.28&  90   & 1.4& 0.74\\
NGC~4522&12 33 40.0 &+09 10 30 & SBcd	  &T/N~[s] & \nodata&-18.11 & 2307& 3.7& 0.27&  33   & 3.3& 0.86\\
NGC~4569&12 36 50.1 &+13 09 48 & SABab    &T/N~[s] & L/Sy   &-20.84 & -235& 9.6& 0.46&  23   & 1.7& 1.47\\
NGC~4579&12 37 44.2 &+11 49 11 & SABb	  &T/N     & L/Sy1.9&-20.62 & 1519& 5.9& 0.79&  95   & 1.8& 0.95\\
NGC~4580&12 37 48.4 &+05 22 09 & SABa pec &T/N~[s] & \nodata&-19.27 & 1034& 2.1& 0.78& 165   & 7.2& 1.53
\enddata
\tablecomments{Col. (1): ID. Col. (2): The right ascension in the
  J2000.0 epoch. Col. (3): The declination in the J2000.0
  epoch. Col. (4): RC3 type. Col. (5): Star formation class of
  \citet{kk04}: N: Normal; A: Anemic; E: Enhanced; T/A:
  Truncated/Anemic;  T/C: Truncated/Compact; T/E: Truncated/Enhanced;
  T/N: Truncated/Normal; T/N~[s] Truncated/Normal (Severe); $^{a}$: classified for this paper. Col. (6):
  Nuclear type: H: H~{\sc ii} region; L: LINER; Sy: Seyfert (1, 1.9,
  2); SB: Starburst. Col. (7): Absolute $B$-band magnitudes. Col. (8):
  Heliocentric velocity. Col. (9): Major axis diameters. Col. (10): 
  Semi-minor/semi-major axis ratio. Col. (11): Position Angle in
  degrees. Col. (12): Distance from Virgo cluster center \citep[$12^{\rm h}30^{\rm m}47^{\rm s}$, $+12\degr 20\arcmin 13\arcsec$;][]{he98} in degrees. Col. (13): Type-independent H~{\sc i} deficiencies, having a typical uncertainty of 0.2~dex arising from the uncertainty in morphological classifications, 
  taken from A. Chung et al. (2008, in preparation) (deficiencies logarithmically increase with number). } 
\end{deluxetable*}

\section{Observations and Data Reduction \label{sec-obs}}

\subsection{Galaxy Sample \label{sec-samp}}
We present {\it Spitzer} far-infrared and VLA radio continuum imagery of 10 Virgo cluster spirals galaxies included in the VLA Imaging of Virgo In Atomic gas (VIVA; A. Chung et al. 2008, in preparation) survey.  
A total of $\sim$40 of the 53 VIVA sample galaxies are included in the {\it Spitzer} Survey of Virgo (SPITSOV; J. D. P.~Kenney et al. 2008, in preparation) imaging program.
The sub-sample of 10 galaxies were the first few observed with {\it Spitzer} that also had high quality radio maps as well as FIR and radio emission detected out about as far as the H~{\sc i} extent.  
Because we are interested in studying the properties of the radio continuum halo to determine if it is interacting with the ICM, we use this criterion to ensure we are detecting infrared and radio continuum emission near the edge of the ISM.  

Basic data for each galaxy can be found in Table \ref{tbl-galdat}. 
Galaxy diameters ($D_{25}$), morphological types, position angles
(PA), and absolute $B$-band magnitudes, assuming a distance of
16.6~Mpc to the Virgo cluster center \citep{shap01}, were taken from
the Third Reference Catalog of Bright Galaxies \citep[RC3;][]{dev91}.   

We also include distances to the Virgo cluster center \citep[$12^{\rm h}30^{\rm m}47^{\rm s}$, $+12\degr 20\arcmin 13\arcsec$;][]{he98}, in degrees, for which the sample ranges between $1.3 - 7.2{\degr}$ (i.e. spanning a factor of $\ga$5).  
Since the Hubble morphological classification appears to have limited
meaning for cluster spirals \citep[e.g. stripped Sc's tend to be
  classified as Sa's;][]{kk98}, we include the H$\alpha$ star
formation type of \citet{kk04}.  
Type-independent H~{\sc i} deficiencies, taken from A. Chung et al. (2008, in preparation), 
are included in Table \ref{tbl-galdat} and have a typical uncertainty of 0.2~dex arising from the uncertainty in morphological classifications.  
The present sample spans a factor of $\sim$50 in H~{\sc i} deficiency
and nearly the entire range in H$\alpha$ star formation type.

To determine if the FIR and radio properties of these cluster galaxies differ systematically from those in the field, we compare with 6 non-Virgo galaxies (NGC~2403, NGC~3031, NGC~3627, NGC~4631, NGC~5194, and NGC~6946), chosen to act as a control sample.  
These galaxies were selected to be less distant than Virgo, span a range in spiral type and inclination, and are a sub-sample of the galaxies included the FIR and radio study of \citet{ejm08}; this investigation describes the physical link between the FIR and radio properties of normal galaxy disks on kiloparsec scales.    
Information on the observations and general characteristics of these galaxies can be found there.  

\subsection{{\it Spitzer} MIPS Imaging}
{\it Spitzer} imaging was carried out for each galaxy using the
Multiband Imaging Photometer for {\it Spitzer} \citep[MIPS;][]{gr04}. 
The strategy of the imaging campaign was based on that of the Spitzer
Infrared Nearby Galaxies Survey \citep[SINGS;][]{rk03} with the only
difference being a factor of 2 increase in exposure times to better
detect diffuse emission arising in the outer regions of the Virgo
sample galaxies.
The MIPS data were reduced as in \citet{jm08} with the following
modifications.  
Since most of the galaxies were observed twice, the data had to be combined into
a single mosaic of the two observations.  
For the 70 and 160$~\micron$ data this was a straightforward process,
combining the filtered basic calibrated data frames (BCDs), which have
near zero residual background, from the two observations using the
Mosaicking and Point Source Extraction (MOPEX) software package.    
The target galaxy was masked during the 70$~\micron$ filtering, as in
\citet{jm08}, while the pipeline filtered BCDs (FBCDs) were used for
160$~\micron$ observations.
The individual 24$~\micron$ observations were first reduced separately
in order to measure the background level.  
The background level was then subtracted from each BCD, and a combined
mosaic was constructed using MOPEX as above.  
Although the two observations for a given galaxy were often performed
in close succession, this was not always the case.  
Measuring and subtracting the two backgrounds was the most reliable
method of placing the two observations on a common scale for
mosaicking. 

For galaxies which were observed once at the time of this writing, 
preliminary results were obtained by combining the BCDs (as above, filtered in the case of 70$~\micron$ and 160$~\micron$) using MOPEX.  
The resulting $160~\mu$m mosaics have small regularly spaced coverage
gaps at the locations of bad pixels on the array. 
This is a consequence of the single redundancy of 160$~\micron$ scan mode
observations, although such occurrences did not affect the field of
view of our galaxies in any observation.
The final calibration uncertainties of 24, 70, and 160~$\micron$ data
are 5, 10, and 15\% respectively.

For galaxies in common with the VIVA survey and SINGS, we use MIPS data included in the SINGS data release 5 (DR5) (see Table \ref{tbl-obsdat}).  
These data were processed using the MIPS Data Analysis Tool \citep[DAT;][]{kdg05} version 3.06.  
Additional steps beyond the standard reduction procedure of the MIPS DAT are described in \citet{ejm06a}.
The full width at half maximum (FWHM) of the MIPS 24, 70, and 160~$\micron$ point spread functions (PSFs) are $\sim$5$\farcs7$, 17$\arcsec$ and 38$\arcsec$ while the final calibration uncertainties
are $\sim$2, 5, and 9\%, respectively.  
The mean RMS noise values of the 24, 70, and 160~$\micron$
data for the SINGS galaxies are $\sim$0.048, 0.45, and
0.57~MJy~sr$^{-1}$ while our deeper imaging campaign reaches mean RMS 
noise values of $\sim$$\sqrt2$ better.  
In some cases (e.g. NGC~4254, NGC~4321, NGC~4579), bright nuclei created negative latent images in the 70~$\micron$ data resulting in bright and dark streaking on opposite sides of the nuclei after background subtraction.    
While visible in some of the images, these streaks are generally low amplitude and should not significantly affect the analysis.  

\subsection{Radio Imaging}
The 1.4~GHz radio continuum maps were created from the line-free
channels of H~{\sc i} data cubes collected as part of of the VIVA
survey; 
a detailed description of the VLA observations along with the H~{\sc i}
reductions and associated data products can be found in A. Chung et al. (2008, in preparation). 
In order to reduce the effects of interfering sources, which causes
substantial sidelobes especially at low frequencies, we have used the
AIPS procedure {\tt PEELR}. 
It iteratively attempts to calibrate on multiple fields around bright
continuum sources (self-calibration), subtract those fields from the 
self-calibrated data, undo the field-specific calibration from the
residual data, and it finally restores all fields to the residual
data. 
The quality of our continuum data varies depending on the number of
line free channels of individual galaxies. 
The same weighting scheme ({\tt ROBUST$=1$}) used for the the H~{\sc i} maps was also used here for the continuum maps.  
The RMS of each 20~cm radio continuum map is given in Table~\ref{tbl-obsdat}.  
The mean RMS noise value of  the VIVA radio maps is $\sim$0.008~MJy~sr$^{-1}$ (0.1~ mJy~bm$^{-1}$) and the range in RMS noise values spans a factor of $\sim$3.  
The FWHM of the CLEAN beam major and minor axes, $B_{\rm maj}$ and $B_{\rm min}$ respectively,  are also given for each galaxy in Table \ref{tbl-obsdat}; the major axes range from $\sim$$17\arcsec-45\arcsec$.

In the case of NGC~4569, a SINGS galaxy observed as part of the Westerbork Synthesis Radio Telescope (WSRT)-SINGS survey \citep{rb07}, we use the WSRT radio map in our analysis since it is an order of magnitude more sensitive than what was achieved by the VIVA survey for this object.  


\begin{deluxetable}{ccccc}
  \tablecaption{Observation Summary\label{tbl-obsdat}}
\tablecolumns{5}
\tablewidth{0pt}
  \tablehead{
    \colhead{} & \colhead{MIPS} & \colhead{$B_{\rm maj}\times B_{\rm min}^{a}$} & 
    \colhead{20~cm RMS}& \colhead{70~$\micron$ RMS}\\
    \colhead{Galaxy} & \colhead{Summary} & \colhead{(arcsecond)} & \colhead{MJy~sr$^{-1}$}&
     \colhead{MJy~sr$^{-1}$}\\
    \colhead{(1)} & \colhead{(2)} & \colhead{(3)} & \colhead{(4)} & \colhead{(5)}}
  \startdata
  \cutinhead{Virgo (Cluster) Sample}
NGC~4254&  $S$&   $34.9\times 23.3$ & 0.0066& 0.45\\
NGC~4321&  $S$&   $27.0\times 22.3$ & 0.0066& 0.24\\
NGC~4330&    1&   $19.4\times 15.4$   & 0.0045& 0.30\\
NGC~4388&    2&   $17.3\times  15.1$ & 0.014& 0.37\\
NGC~4396&    2&   $20.5\times 20.2$  & 0.0047& 0.18\\
NGC~4402&    1&   $17.1\times  15.2$  & 0.0097& 0.54\\
NGC~4522&    2&   $19.0\times   15.3$ & 0.0066& 0.26\\
NGC~4569$^{b}$&  $S$&  $45.0\times 45.0$ & 0.0015& 0.22\\
NGC~4579&  $S$&   $43.2\times 35.2$& 0.0048& 0.26\\
NGC~4580&    1&   $17.5\times 16.6$& 0.0094 & 0.50\\
\cutinhead{Field (Control) Sample$^{c}$}
NGC~2403& $S$& $92.7\times 92.7$& 0.0020& 0.13\\
NGC~3031& $S$& $82.2\times 82.2$& 0.0028& 0.13\\
NGC~3627& $S$& $31.8\times 31.8$& 0.0039& 0.45\\
NGC~4631& $S$& $39.0\times 39.0$& 0.0023& 0.24\\
NGC~5194& $S$& $38.4\times 38.4$& 0.0015& 0.24\\
NGC~6946& $S$& $43.9\times 43.9$& 0.0036& 0.35
\enddata
\tablecomments{Col. (1): ID. Col. (2): {\it Spitzer} MIPS Data: $S$:
  Observed as part of SINGS; 1: Only a single AOR included;
  2: Both AORs included. Col. (3): Major and minor axes of the original radio
  CLEAN beam used in the VIVA radio continuum maps (A. Chung et al. 2008, in preparation).
  Col. (4): 20~cm RMS. Col. (5): 70~$\micron$ RMS.\\
  $^{a}$: The FWHM of the PSF at 70~$\micron$ is $\sim$17$\arcsec$ for each galaxy; $B_{\rm maj}$ therefore sets the resolution for the analysis of each galaxy disk.\\    
  $^{b}$: For NGC~4569 we use the WSRT-SINGS radio continuum map of \citet{rb07} rather than the VIVA radio continuum data because the WSRT-SINGS map is significantly more sensitive.\\
  $^{c}$: These data were taken from \citet{ejm08}; the resolution indicates that for which the galaxy's maps were smoothed to simulate their appearance at the distance of Virgo.  It is at this resolution for which the RMS noise was measured.
}
\end{deluxetable}

\subsection{Image Registration and Resolution Matching
  \label{sec-resmatch}}  
To perform an accurate comparative analysis between the MIPS and radio data we match the resolution of the final calibrated images using the MIPS PSF. 
After cropping each set of galaxy images to a common field of view we CLEANed the radio data and convolved the resulting CLEAN components with a model of the MIPS 70~$\micron$ PSF.  
We restore with the MIPS beam first because it has a non-Gaussian shape with significant side-lobes.  

Since the FWHM of the original radio CLEAN beam $B_{\rm maj}$ was in all cases  larger than the FWHM of the MIPS 70~$\micron$ PSF, we convolved the 70~$\micron$ and 20~cm data with an additional Gaussian beam to account for this difference.       
These images were then re-gridded to a common pixel scale and properly aligned. 
The observed 70~$\micron$ and resolution-matched 20~cm maps for each galaxy are presented in the first and second columns of Figure \ref{fig-maps}.
The effective resolution used in the comparison of the infrared and radio data is given in Table \ref{tbl-obsdat}.

\subsection{Global Flux Densities and FIR/Radio Ratios
  \label{sec-flux}} 
In Table \ref{tbl-flux} we give the global MIPS and 20~cm flux
densities.
We include extended source aperture corrections for the MIPS fluxes by
adopting the median values derived by \citet{dd07} which are 1.01,
1.04, and 1.10 at 24, 70, and 160~$\micron$ respectively.
Because our sample does not include dwarf galaxies, for which the
corrections were found to be large, simply taking the median value
should be adequate.
Since we are using the WSRT-SINGS radio continuum map of NGC~4569, we also work with the global flux density at 22~cm reported by \citet{rb07} which has been scaled to what is expected at 20~cm assuming a synchrotron power law of the form $S_{\nu} \propto \nu^{-\alpha}$ with a radio spectral index $\alpha = 0.8$ \citep[i.e. typical for star-forming galaxies;][]{nkw97}.   
This was also done for the 22~cm flux densities of the WSRT-SINGS field galaxies where appropriate.  

We also provide measures of the total infrared flux (TIR;
$3-1100~\micron$) along with logarithmic far-infrared (FIR;
$42-122~\micron$)-to-radio ratios in the form of the commonly used
parameter $q$ \citep{gxh85} such that, 
\begin{equation}
\label{eq-qFIR}
q \equiv \log~\left(\frac{F_{\rm FIR}}{3.75\times10^{12}
  ~{\rm W m^{-2}}}\right) - \log~\left(\frac{S_{1.4~{\rm GHz}}}{\rm W
  m^{-2} Hz^{-1}}\right).
\end{equation}
The TIR fluxes were measured using a linear combination of the 3 MIPS
bands given by Equation 4 of \citet{dd02}. 
The corresponding FIR fractions were then calculated using the same
spectral energy distribution (SED) models of \citet{dd02} as given by
Equation 3 of \citet{ejm08}, i.e., 
\begin{equation}
\label{eq-firfrac}
\frac{F_{\rm FIR}}{F_{\rm TIR}} = \displaystyle\sum_{i=0}^{3}
\xi_{i}\log~\left(\frac{f_{\nu}(70~\micron)}{f_{\nu}(160~\micron)}\right)^{i}, 
\end{equation}
where \([\xi_{0},\xi_{1},\xi_{2},\xi_{3}] = [0.5158, 0.1619, -0.3158,
  -0.1418]\) and \(-0.65 \leq
\log~f_{\nu}(70~\micron)/f_{\nu}(160~\micron) \leq 0.54\).  
  
\begin{deluxetable*}{ccccccc}
  \tablecaption{Global Flux Densities and Derived Parameters
  \label{tbl-flux}} 
  \tablecolumns{7}
  \tablehead{
    \colhead{} &  \colhead{$S_{\nu}$(20~cm)} &
     \colhead{$f_{\nu}$(24~$\micron$)}  &
     \colhead{$f_{\nu}$(70~$\micron$)}  &
     \colhead{$f_{\nu}$(160~$\micron$)} &
     \colhead{$F_{\rm TIR}$} & \colhead{}\\
     \colhead{Galaxy} & \colhead{(Jy)} & \colhead{(Jy)} &
     \colhead{(Jy)} & \colhead{(Jy)} &
     \colhead{(10$^{-13}$~W~m$^{-2}$)} &
     \colhead{$q$}
  }
  \startdata
  \cutinhead{Virgo (Cluster) Sample}
NGC~4254&  0.439&    4.17&   45.0&    147 &     60.0&   2.13\\
NGC~4321&  0.273&    3.44&   36.8&    143 &     54.9&   2.26\\
NGC~4330&  0.020&    0.107&    1.05&      5.65&      1.97&   1.92\\
NGC~4388&  0.164&    2.64&   10.6&     20.0&     13.7&   2.00\\
NGC~4396&  0.022&    0.132&    1.74&      5.79&      2.29&   2.00\\
NGC~4402&  0.067&    0.689&    7.52&     25.5&     10.2&   2.17\\
NGC~4522&  0.027&    0.217&    1.43&      6.60&      2.55&   1.91\\
NGC~4569&0.157$^{a}$&1.41&   11.2&     41.5&     16.8&   2.00\\
NGC~4579&  0.167&    0.816&    9.34&     42.8&     15.4&   1.89\\
NGC~4580&  0.005&    0.148&    1.83&      7.75&      2.84&   2.73\\
\cutinhead{Field (Control) Sample$^{b}$}
NGC~2403 &0.33$^{a}$    & 5.84 &   86.4 &  246   &  102  &	 2.50 \\
NGC~3031 &0.55$^{a}$    & 5.09 &   85.2 &  360   &  129  &	 2.31 \\
NGC~3627 &0.46$^{a}$    & 7.42 &   92.6 &  230   &  103  &	 2.38\\
NGC~4631 &1.20$^{a}$    & 8.15 &  130   &  290   &  132  &	 2.09\\
NGC~5194 &1.15$^{a,c}$&12.7 &  147   &  495   &  198  &	 2.22\\
NGC~6946 &1.57$^{a}$   & 20.4&  207   &  503   &  234  &	 2.21
\enddata
\tablecomments{
$^{a}$: Scaled from the 22~cm flux density reported by \citet{rb07} assuming a radio spectral index $\alpha = 0.8$ (see $\S$\ref{sec-flux}).\\
$^{b}$: Data taken from Table 2 of \citet{ejm08}, where MIPS flux densities were taken from \citet{dd07}.  $q$ values have been adjusted since here we are using 20~cm radio flux densities rather than 22~cm flux densities.\\
$^{c}$: Corrected for flux contributions from companion galaxy NGC~5195.}
\end{deluxetable*}

\begin{figure*}[!ht]
  \plotone{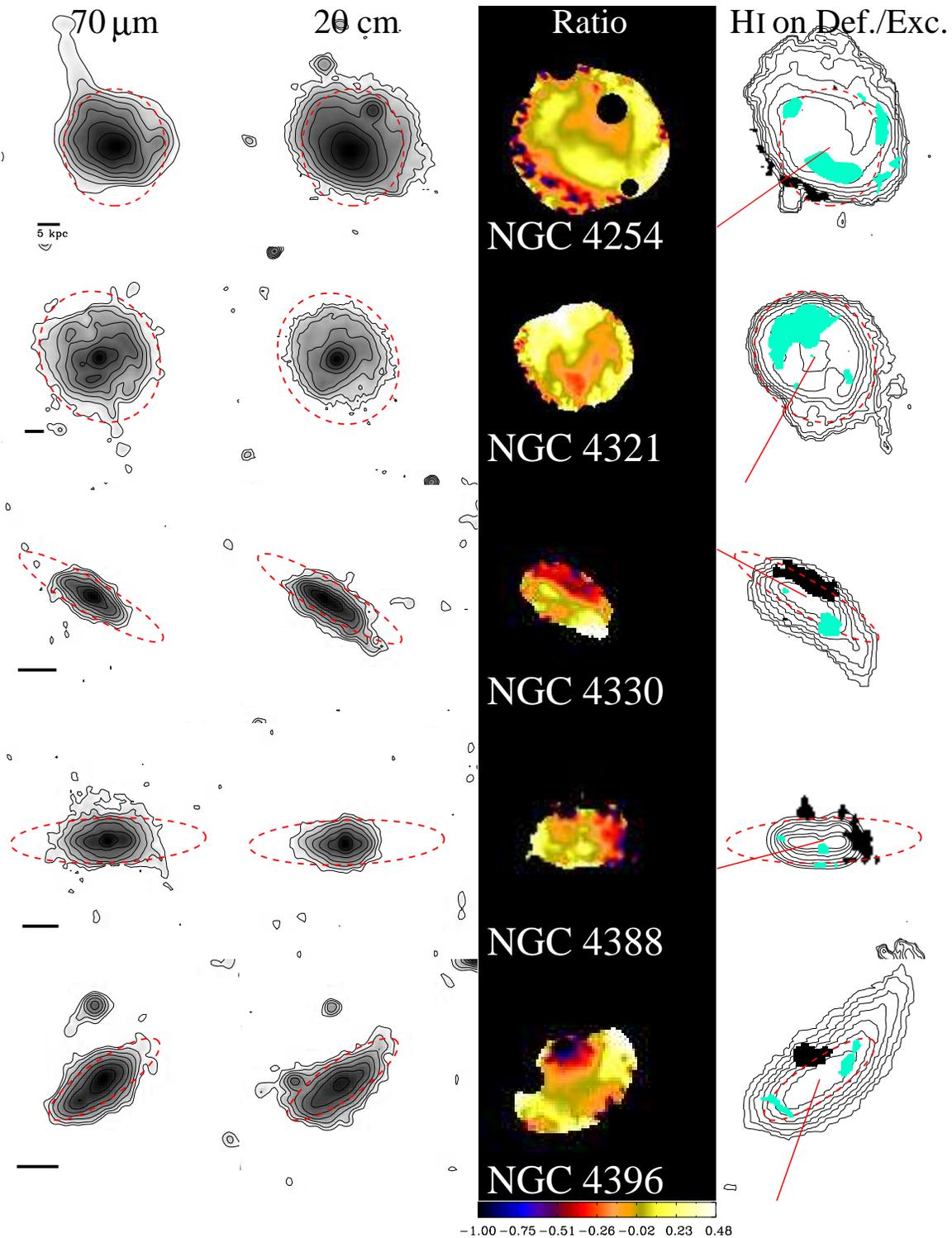}
\caption{
  Each galaxy's 70~$\micron$ and 22~cm images are displayed in columns
  1 and 2, respectively.
  Streaking artifacts in the 70~$\micron$ data of NGC~4254, NGC~4321,
  and NGC~4579 are easily identifiable.
  The stretch of gray-scale and contours ranges, logarithmically, from the 3-$\sigma$ RMS level of
  the background to the maximum surface brightness of the galaxy. 
  In column 3 the logarithm of the ratio of observed to predicted
  radio flux (see $\S$\ref{sec-modrc}) is shown with a stretch ranging
  from 10\% (dark) to a factor of 3 (light). 
  The radio deficit and excess regions (see $\S$\ref{sec-rcdef-def}) are given in
  column 4 along with a vector showing the direction to cluster center.
  Overlaid on each deficit (black) and excess (cyan) region map are H~{\sc i} contours;  
  the H~{\sc i} contours begin at the $\sim$3-$\sigma$ RMS level
  and increase logarithmically.
  Using a dashed line we overlay the extent of the optical disk on the 70~$\micron$, 22~cm, and H~{\sc i} maps.   
  A 5~kpc scale bar is also drawn in each of the 70~$\micron$ image panels, assuming a distance of 16.6~Mpc to the Virgo cluster center.
  \label{fig-maps}}
\end{figure*}

\setcounter{figure}{0}
\begin{figure*}[!ht]
\plotone{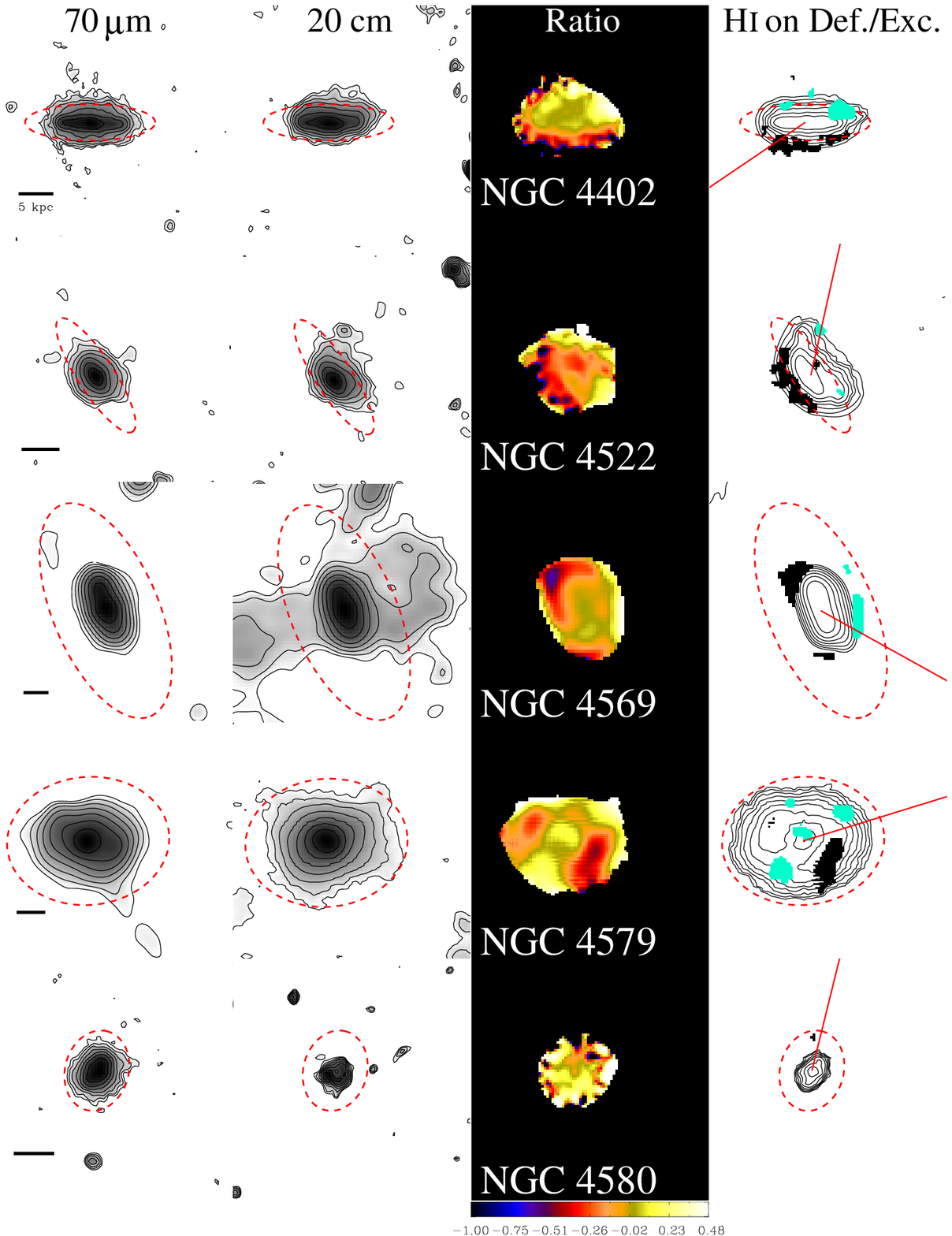}
\caption{\it Continued}
\end{figure*}

\begin{deluxetable}{ccc}
  \tablecaption{Image-Smearing Parameters\label{tbl-scl}}
 \tablecolumns{3}
 \tablehead{
    \colhead{} & \colhead{$l$} & \colhead{$\Phi$}\\
    \colhead{Galaxy} & \colhead{(kpc)} & \colhead{(dex)}}
  \startdata

\cutinhead{Virgo (Cluster) Sample}
NGC~4254&  2.3&  0.927\\
NGC~4321&  0.7&  0.253\\
NGC~4330&  0.7&  0.381\\
NGC~4388&  0.2&  0.078\\
NGC~4396&  0.9&  0.668\\
NGC~4402&  0.6&  0.581\\
NGC~4522&  0.8&  0.835\\
NGC~4569&  0.8&  0.933\\
NGC~4579&  0.0&  0.000\\
NGC~4580&  0.8&  0.762\\

\cutinhead{Field (Control) Sample$^{a}$}
NGC~2403& 1.0& 0.664\\
NGC~3031& 1.8& 0.464\\
NGC~3627& 0.5& 0.541\\
NGC~4631& 0.6& 0.768\\
NGC~5194& 0.5& 0.343\\
NGC~6946& 0.6& 0.590
\enddata
\tablecomments{$^{a}$: Values taken from \citet{ejm08}.  
}
\end{deluxetable}

\begin{figure}[!ht]
  \plotone{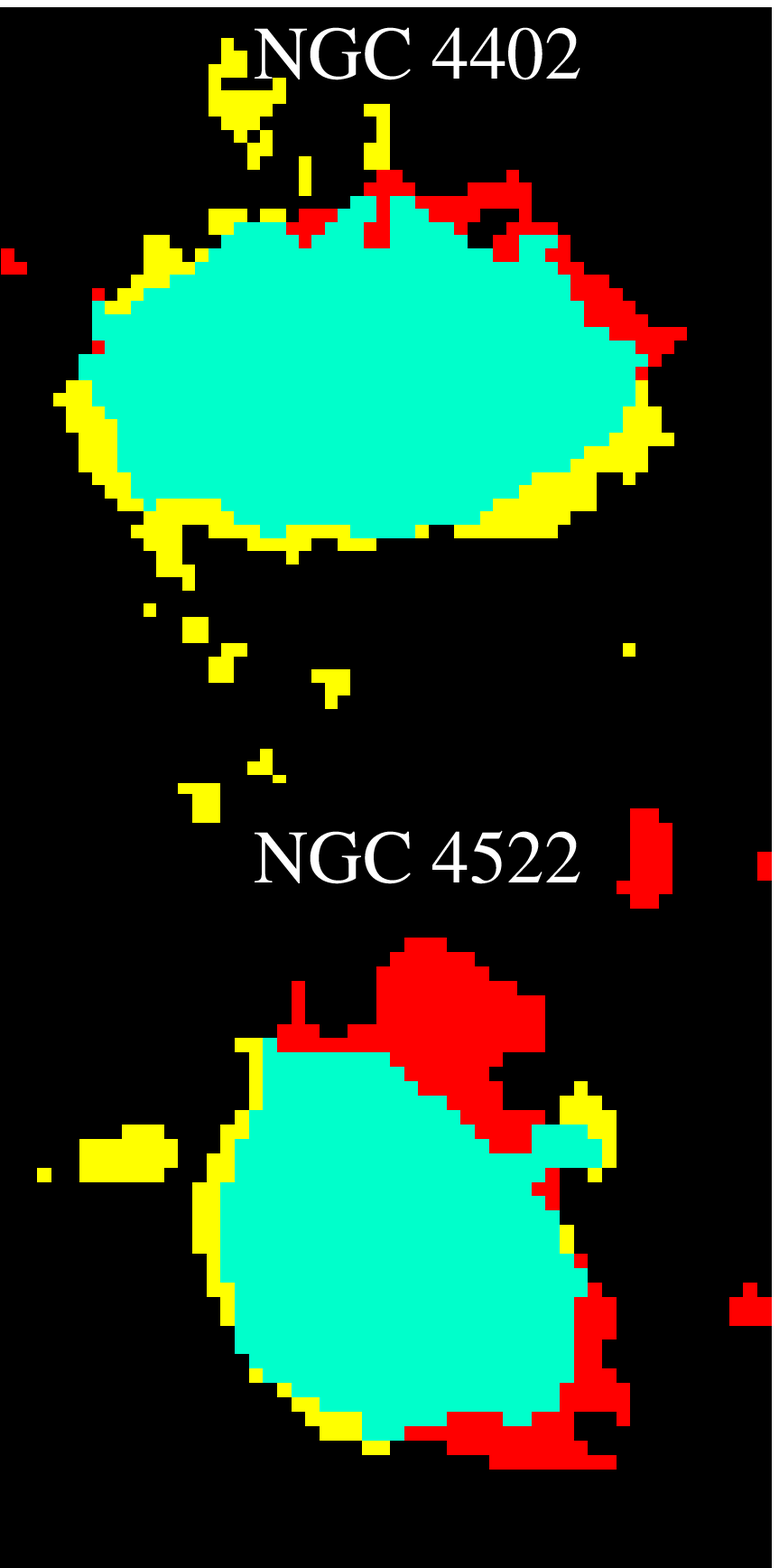}
\caption{Pixel identification maps indicating regions of the galaxy
  for which detections were made at the 3-$\sigma$ RMS level.
  Cyan regions indicate areas where both the radio and 70~$\micron$
  emission were detected at the 3-$\sigma$ RMS level while red and
  yellow regions identify where only the radio or 70~$\micron$
  emission has been detected at the 3-$\sigma$ RMS
  level.  \label{fig-pixid} 
}  
\end{figure}

\begin{figure*}[!ht]
  \plotone{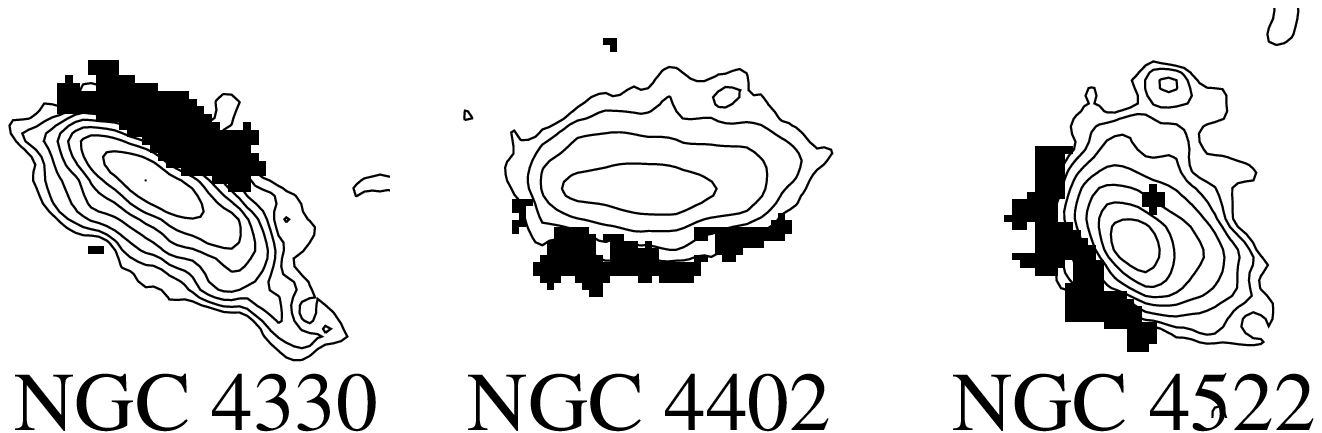}
\caption{
The radio deficit regions of NGC~4330, NGC~4402, and NGC~4522 with radio continuum contours.  The radio continuum contours begin at the 3-$\sigma$ RMS level and increase logarithmically. \label{fig-tails}}  
\end{figure*}

\section{Identifying ICM Related Effects in the Radio Morphologies
  \label{sec-analysis}} 
\subsection{Radio and Infrared Morphologies}
A comparison of the 70~$\micron$ and 20~cm images in the first two columns of Figure \ref{fig-maps} shows that the radio continuum images for some of the galaxies are quite disturbed, appearing asymmetric compared to the infrared images (e.g. NGC~4330 and NGC~4522).  
In some cases we find regions along edges having FIR emission without corresponding radio emission (these are discussed further in $\S$6.4); 
regions opposite of these edges often have radio continuum tails without FIR counterparts.  
To illustrate this we plot pixel identification maps for NGC~4402 and NGC~4522 in Figure \ref{fig-pixid}; regions of cyan indicate that both the radio and 70~$\micron$ emission was detected at the 3-$\sigma$ RMS level while region of red and yellow indicate areas where only the radio or only the 70~$\micron$ emission has been detected at the 3-$\sigma$ RMS level.  

The FIR surface brightness depends on the ISM density and heating intensity by starlight.
Assuming that the FIR dust emission is, to first order, tracing the distribution of molecular gas within galaxies \citep[e.g.][]{mm05}, the observed differences in the radio continuum and FIR emission distributions suggest that the relativistic ISM is perturbed in cluster galaxies more easily than molecular material.  
We would like now to interpret and quantify these effects.  

\subsection{Modeled Radio Continuum Maps \label{sec-modrc}}
Detailed studies of the FIR-radio correlation within nearby field
galaxies \citep[e.g.][]{ejm06a, ejm08} have shown that the dispersion
in the FIR/radio ratios on $\ga$$0.1 - 1$~kpc scales within galaxy
disks is similar to the dispersion in the global FIR-radio correlation
(i.e. $\la$0.3~dex).  
This strong correlation between the FIR and radio images of spirals has
been found to improve by a factor of $\sim$2 using an image-smearing
technique which approximates a galaxy radio image as a smoother version
of its FIR image due to the diffusion of CR electrons.
While the star-forming disks of cluster galaxies are often truncated due to gas stripping events, the images of the remaining 70~$\micron$ disks for the entire sample do not appear to be otherwise more strongly disturbed than galaxies in the field; 
we do not observe any obvious sharp edges, tails, or evidence for strong extraplanar emission at 70~$\micron$.  
Therefore, we apply the image-smearing technique of \citet{ejm08}
(using a single smoothing function) to create models for the expected
radio distribution of a galaxy assuming that the FIR/radio ratio map is like that
of a field galaxy; the modeled radio continuum images are denoted as $R_{\rm mod}$.  

Since the image-smearing description is applicable purely for
non-thermal radio emission, we include an estimate of the
thermal radio emission $R_{\rm T}$ derived from the 24~$\micron$ flux density
distribution \citep[see][for details]{ejm06b,ejm08} such that,
\begin{equation} 
\label{eq-rctherm}
\left(\frac{R_{\rm T}}{\rm Jy}\right) \sim 7.93\times10^{-3}
\left(\frac{T}{10^{4}~{\rm K}}\right)^{0.45} 
\left(\frac{\nu}{\rm GHz}\right)^{-0.1}
\left(\frac{f_{\nu}(24~\micron)}{\rm Jy}\right), 
\end{equation}
where we have assumed an average H~{\sc ii} region temperature of $T = 10^{4}$~K.

We compare the observed radio continuum images with the model
radio continuum images, $R_{\rm mod}$, such that    
\begin{equation} 
\label{eq-modrc}
R_{\rm mod} = R_{\rm T} + Q_{\rm NT}^{-1}(\kappa(l) \ast I),
\end{equation}
where $R_{\rm T}$ is an estimate of the thermal radio emission
distribution using Equation \ref{eq-rctherm}, $I$ is the 70~$\micron$ image, 
$\ast$ denotes a convolution with an exponential kernel $\kappa$ having scale-length $l$, 
and $Q_{\rm NT}$ is the global $f_{\nu}(70~\micron)/S_{\nu}(20~{\rm cm})$ flux density
ratio after correcting the global 20~cm flux density for thermal radio emission using the galaxy's 24~$\micron$ flux density and Equation \ref{eq-rctherm}.   
The scale-length $l$ is determined by minimizing the residuals
between the radio and smoothed infrared maps as described by
\citep{ejm06a,ejm08};
only areas of the galaxy disk which are detected above the 3-$\sigma$
RMS noise in both of the observed radio and infrared images are
considered.
Thus, the residuals are calculated over the non-disturbed regions of
the radio disks.     

To quantify the improvement in the spatial correlation between
the radio and smoothed 70~$\micron$ images we measure $\Phi$ which is
the logarithmic decrement from the squared sum of the residuals between the
{\it observed} (unsmoothed) radio and infrared images to the minimum residuals which defines the best-fit scale-length.  
The values of the best-fit scale-lengths used for creating the modeled
radio maps along with the improvement in the residuals between the two
maps, as defined by $\Phi$, are given in Table \ref{tbl-scl}.

We find that the the image-smearing improves the dispersion between
the infrared and radio maps by a factor of $\sim$2, on average; 
this is the same quantitative improvement quoted by \citet{ejm08}
using largely field galaxies.   
We also find best-fit scale-lengths of 0.2~kpc for NGC~4388 and 0.0~kpc for NGC~4579, which likely result from their bright Seyfert-type nuclei.

\subsection{Deviations from Expectations for Field Galaxies
  \label{sec-rcdef-def}} 
Our aim is to determine if differences between the observed and
modeled radio continuum images arise from ICM-ISM interactions and, if
so, whether quantifying and comparing such differences within our
sample can help improve our current physical picture of such
interactions.  
We therefore create ratio maps between the observed and modeled radio
maps in the following  manner.
We divide the observed radio map by the modeled radio map, keeping only those pixels ($r_{i}$) detected at $\geq$5-$\sigma$ RMS level of either map.
We also construct a map giving the associated uncertainty at each
pixel of the ratio map ($\sigma_{r_{i},{\rm RMS}}$) by numerically
propagating the uncertainties of the input images as measured by the
1-$\sigma$ RMS of their maps.  
Any background radio galaxies or artifacts arising from streaking in
the 70~$\micron$ data (e.g. NGC~4254, NGC~4321, and NGC~4579) outside
the galaxy disk are removed.  
The ratio maps are displayed in column 3 of Figure \ref{fig-maps}.

Next, we identify regions for which the ratio of observed to modeled radio flux density departs significantly from unity.
To do this we compute the interquartile standard deviation of the ratio map which is the standard deviation computed after removing values included in the first and fourth quartiles;
we denote this value as $\sigma_{r,{\rm disk}}$ which, for a Gaussian distribution, is equal to $\sim$$\onethird$ the population standard deviation. 
Radio deficit regions are constructed using pixels which
simultaneously obey the following two conditions: 
$\{|1 - r_{i}|/\sigma_{r_{i},{\rm RMS}} > 2.5\}$ and $\{(1 -
r_{i})/\sigma_{r,{\rm disk}} > 2.5\}$.
The first condition selects
pixels which are significantly deviant with respect to the RMS noise of the maps while the second condition selects pixels which are significantly deviant with respect to the intrinsic dispersion of the ratio map within the galaxy disk.  
Similarly, radio excess regions are constructed from all those pixels in the ratio map which simultaneously obey the conditions such that: $\{|1 - r_{i}|/\sigma_{r_{i},{\rm RMS}} > 2.5\}$ and $\{(r_{i} - 1)/\sigma_{r,{\rm disk}} > 2.5\}$.  

Radio excesses appear largely coincident with regions associated with synchrotron tails like those observed in NGC~4330, NGC~4402 and NGC~4522. 
In the cases of NGC~4524, NGC~4321, and NGC~4579, the radio excess regions appear to be possibly associated with the streaking in the 70~$\micron$ data.  
On the other hand, for most of our sample galaxies we detect radio deficit regions which are 
located along a single edge of their disks.  
The radio deficit regions are generally found opposite radio excess regions that are associated with synchrotron tails; 
this is illustrated in Figure \ref{fig-tails} for NGC~4330, NGC~4402, and NGC~4522.    
NGC~4254 is an exception, for which an excess region is also found interior to the deficit region.
Since these synchrotron tails are likely identifying the direction of the ICM wind, we focus our attention toward the radio deficit regions as they are probing directly the most intense effects of the ongoing ICM-ISM interactions.   

We define the maximum ratio, or fractional deficit, used for identifying radio deficient pixels as $r_{\rm cut}^{\rm def} = 1 - 2.5\sigma_{r,{\rm disk}}$ and find $<r_{\rm cut}^{\rm def}> \sim 0.5 \pm 0.1$ among the sample galaxies.
We note that the lack of high-quality radio continuum data for the entire sample is a limiting factor for this study.  
The large range in the RMS noise of the radio maps (see Table \ref{tbl-obsdat})
prohibits a proper comparison between the deficit regions for each galaxy.  
However, since there is such small scatter in $r_{\rm cut}^{\rm def}$, we choose to use a constant value of $r_{\rm cut}^{\rm def} = 0.5$, or 50\% of the expected flux, for creating the deficit region maps of all galaxies; this allows for the most reliable inter-galaxy comparison.  
Doing the same for excess regions we find $<r_{\rm cut}^{\rm exc}> \sim 1.3 \pm 0.1$ and choose a constant value of $r_{\rm cut}^{\rm def} =1.3$, or 1.3 times the expected flux, for creating the
excess region maps of all galaxies.
The radio continuum deficit (black) and excess (cyan) regions are given in column 4 of Figure \ref{fig-maps} and are overlaid by the contours of the VIVA H~{\sc i} images. 

To compare quantitatively the radio deficiencies among the sample
galaxies we define two parameters.
First, we measure the ratio of the deficit region area ($A_{\rm def}$) to that of the total galaxy area given by the size of our ratio maps ($A_{\rm disk}$) which are determined by the regions of either significant FIR or radio emission.  
Next, we calculate the severity of the deficit as measured by the parameter 
\begin{equation}
\Upsilon = \frac{(S_{\nu}^{\rm mod} - S_{\nu}^{\rm obs})_{\rm
    def}}{S_{\nu}^{\rm glob}}
\end{equation}
which measures the difference between the observed and modeled
radio flux densities within the radio deficit region normalized by
the global radio flux density of the galaxy (i.e. the flux densities
given in Table \ref{tbl-flux}).  
These values along with the flux densities and areas of the deficit
regions are given in Table \ref{tbl-rcdefres}.

\section{Results Comparing Observed and Predicted Radio Distributions
  \label{sec-results}} 

\subsection{The Ratio Maps \label{sec-ratmaps}}
The ratio maps in column 3 of Figure \ref{fig-maps} exhibit gradients near one edge of the galaxy disks for most of the sample;
the lowest ratios lie preferentially along these edges.   
A comparison of these gradients with our beam demonstrates that they are in fact resolved.  
Furthermore, these gradients are opposite to what is expected if we were not smoothing the infrared images enough to estimate properly the CR-electron diffusion lengths since, as observed, the radio disks of galaxies are generally found to be characterized by a larger scale-length than their infrared disks \citep[e.g.][]{bhc89,bh90,mh95}.   
Unlike the rest of the sample, we do not identify statistically significant radio deficiencies within the disks of NGC~4321 and NGC~4580.

We do detect radio deficit regions in NGC~4396 and NGC~4579, however
these regions may arise from non-ICM related effects.
The radio deficit region for NGC~4396 is detected to the north next to a number of nearby background galaxies.  
It is unclear if the observed radio deficit region partly due to a confusion artifact resulting from the background galaxies,  or if it is real.  
Unlike the infrared map, the radio continuum and H~{\sc i} images
appear to have a sharp edge coincident with the radio deficit region; 
however, this region is not opposite the location of extended H~{\sc i} emission.  
In the case of NGC~4579 we find a radio deficit region interior to the
edge of the H~{\sc i} disk.  
The region is consistent with a streaking artifact in the 70~$\micron$
image resulting from the galaxy's bright, Seyfert-type nucleus.

For the remaining six galaxies (i.e. 60\% of the sample), regions
identified to be radio deficient are always at an edge of the disk and
opposite of any identified H~{\sc i} tails.  
For example, H~{\sc i} tails are clearly identified by the contour overlay on the radio continuum deficit region maps of NGC~4330 and NGC~4522 in column 4 of Figure \ref{fig-maps}.    
H~{\sc i} tails are also observed for, and opposite to the radio deficit regions of NGC~4254, NGC~4388, NGC~4402, and NGC~4569, however they are not easily visible in Figure \ref{fig-maps} either because the VIVA data is not sensitive enough to pick up the low level emission (i.e. NGC~45254 and NGC~4388) or because of smoothing the data for resolution matching (i.e. NGC~4402 and NGC~4569).  
The radio continuum deficit regions also appear opposite any observed radio continuum tails which is clearly illustrated 
for NGC~4330, NGC~4402, and NGC~4522 in Figure \ref{fig-tails}.
The combination of these observations suggests that the radio deficit regions likely arise from the same gravitational or gasdynamical effects which have displaced the galaxy's H~{\sc i} gas and relativistic plasma.  
Since we think that the H~{\sc i} tails are created by the ICM wind, we therefore believe that the radio deficit regions identify the zone in which the ICM wind is actively working on each galaxy's ISM.  

For those sample galaxies with radio deficit regions, the fractional
area of the disk defined to be radio deficient ranges from $\sim$5 to 30\%.  
We also find that the difference between the observed and modeled flux densities within the deficit region, normalized by the global flux density (i.e the parameter $\Upsilon$), range from $\sim$1 to 15\%.
Not surprisingly, we find that deficit area ratio increases with increasing $\Upsilon$ (see Table \ref{tbl-rcdefres}).

\subsection{Comparison with Field Galaxies \label{sec-singscomp}}
To determine whether the identification of the radio continuum deficit regions is unique to these cluster galaxies we repeat the above analysis on six non-Virgo galaxies observed as part of SINGS:
NGC~2403, NGC~3031, NGC~3627, NGC~4631, NGC~5194, and NGC~6946.  
A detailed comparison of the FIR and radio properties of these galaxies, along with the rest of the WSRT-SINGS sample, can be found in \citet{ejm08}.     
We note that \citet{ejm08} do not report finding asymmetric gradients in the FIR/radio ratio maps for their sample of mostly field galaxies, suggesting that the FIR and radio properties of these Virgo galaxies do differ systematically from those in the field.  
To test this further, we repeat the exercise of looking for radio deficit regions among these six galaxies.  

\begin{figure*}[!ht]
  \plotone{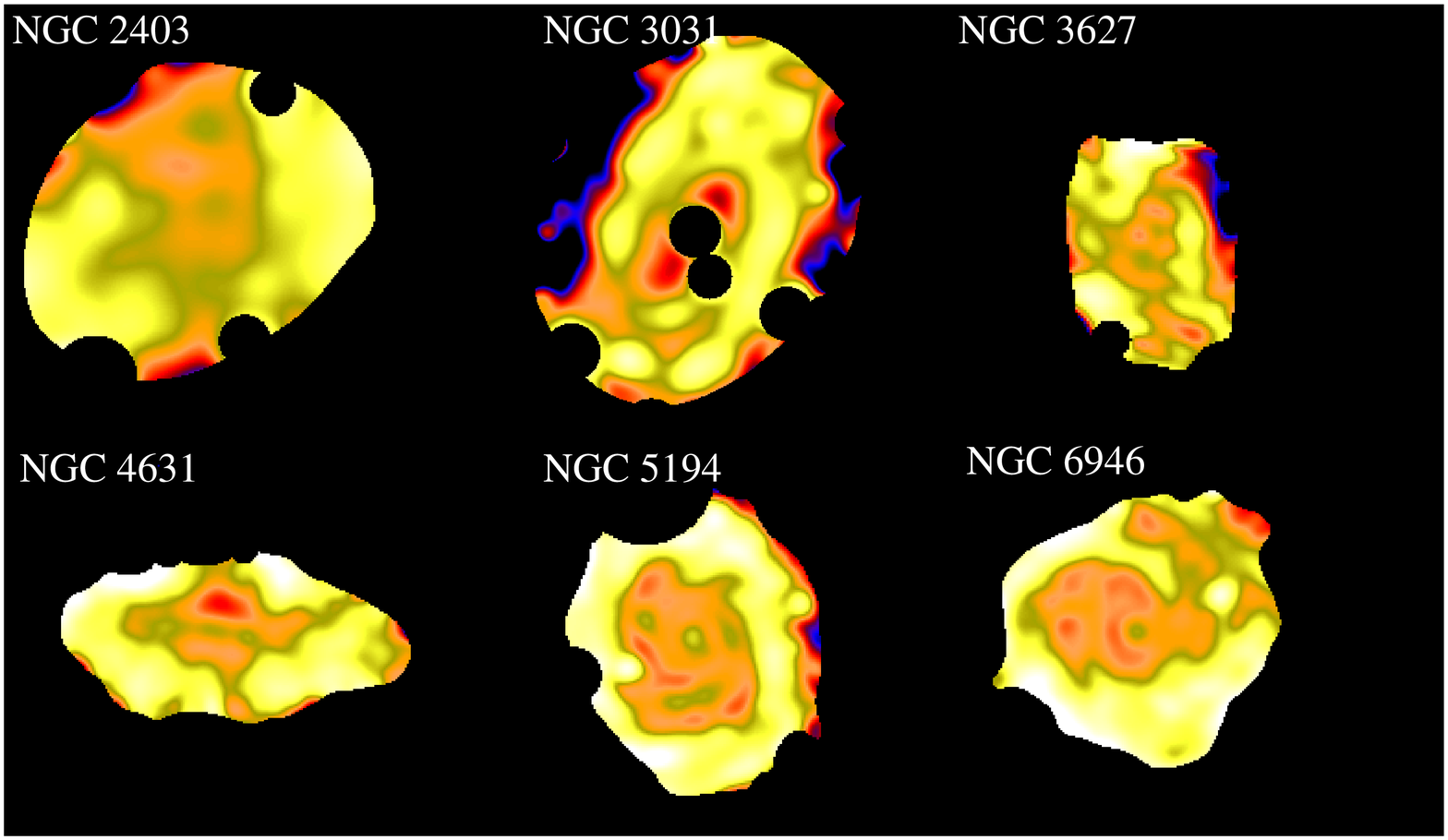}
\caption{Ratio maps of the observed to expected radio flux for six
  galaxies observed as part of SINGS (see $\S$\ref{sec-rcdef-def}).
  The input radio and infrared images were smoothed to simulate the
  expected appearance of each galaxy at the distance of Virgo.
\label{fig-singsmaps}}
\end{figure*}

The smoothing scale-lengths used to create the modeled radio maps are given in Table \ref{tbl-scl}.  
However, before constructing model radio and ratio maps, the infrared and radio images of these galaxies were smoothed appropriately to simulate their appearance at the distance of Virgo.  
This ensures distance effects, such as differences in the physical resolution of each disk, will not introduce any biasses into the comparison.  
The final beam sizes, along with the corresponding noise in the radio and infrared maps, are given in Table \ref{tbl-obsdat}.       
The ratio maps of each galaxy are given in Figure \ref{fig-singsmaps} using the same color scale as for the ratio maps of the Virgo galaxies in the third column of Figure \ref{fig-maps}.

Unlike some of the ratio maps for the Virgo sample (e.g. NGC~4402 and NGC~4522) we do not find strong gradients near the edges of the disks for the non-Virgo galaxies.     
We also note that it was not necessary to show maps of radio continuum deficit regions for the WSRT-SINGS galaxies  
since, except for two cases, radio deficits were not detected.  
The first exception is NGC~3031 where we find a few pixels (i.e. $\la$0.8\% of the
total area) to be radio deficient just surrounding its nucleus; this likely arises from its bright AGN.  
The other exception is NGC~3627, in which the north-west region of the galaxy is significantly radio deficient (i.e. $\Upsilon =0.011$ and $A_{\rm def}/A_{\rm disk} =0.08$ as noted in Table \ref{tbl-rcdefres}); among the Virgo galaxies with radio deficit regions detected, only NGC~4524 has a lower $\Upsilon$ value.   
The radio deficit region is located opposite of its observed H~{\sc i} tail \citep{mh79} and just north of an area observed to have highly polarized radio emission \citep{ms99,ms01}.
The consistency in the location of the deficit region with other
tracers for ongoing ICM-ISM interactions, combined with the fact 
that NGC~3627 is a member of the Leo Triplet, suggests that the radio
deficit region may arise from an interaction within the group.  
The control sample behavior clearly demonstrates that the radio deficiency morphology observed in Virgo galaxies is singular and most naturally explained as a result of ICM-ISM interactions.   

\section{NGC~4522: A Well Studied Case of Ram Pressure Stripping
  \label{sec-n4522}} 

\subsection{Atomic and Ionized Gas}
The Virgo spiral NGC~4522 is one of the clearest cases where the effects of
ram pressure, resulting from the galaxy's rapid motion through the
ICM, are directly observed \citep[e.g.][]{kk99,kvv04,bv04,hc06}. 
A combination of observational data including asymmetric and
extraplanar emission in the H$\alpha$ \citep{kk99} and H~{\sc i}
\citep{kvv04} lines are highly suggestive of such ICM-ISM
interactions.  
While the ICM wind has stripped away atomic gas, apparent by the
extraplanar H~{\sc i} gas, the extraplanar H$\alpha$ emission is
thought to arise from extraplanar H~{\sc ii} regions;
so it appears that virtually the entire ISM has been stripped away
from the outer disk, including molecular clouds.
The stripping appears to have occurred very recently.
Both the stellar population study of the stellar disk beyond the gas
truncation radius \citep{hc06} and a comparison of the observed gas
distribution and kinematics with simulations \citep{bv06} indicate
that the galaxy was stripped within the last $50-100$~Myr.
Thus, it is likely that the current ram pressure is still
significant.  

The H~{\sc i} contour overlay on the radio deficit region
of NGC~4522 in Figure \ref{fig-maps} shows that the deficit region
runs along the eastern edge of the galaxy, opposite the
extraplanar H~{\sc i} gas.
Furthermore, the radio continuum emission itself appears asymmetric;
the radio disk has a sharp cutoff along the eastern edge and
extraplanar emission past the western edge similar to that of the 
H~{\sc i} gas.  
This is consistent with the picture in which the ICM is acting on
the eastern (leading) edge of the galaxy disk, where we find the radio
continuum to be deficient, and is pushing the ISM (gaseous and
relativistic phases) toward and beyond the western edge.

\subsection{Radio Polarization and Spectral Index Maps}
\begin{figure}
\plotone{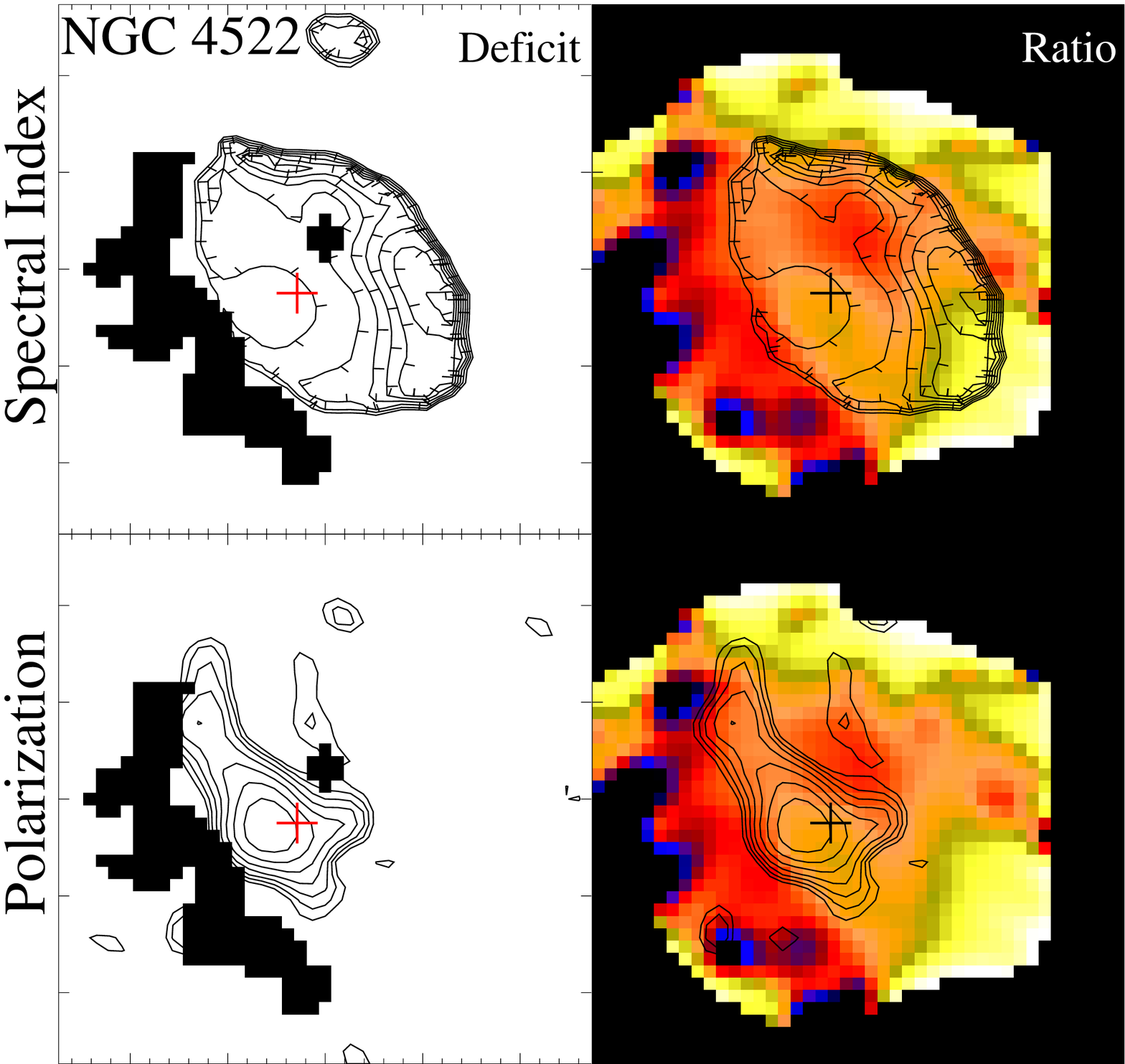}
\caption{ 
  In the top panels, from left to right, we plot the radio continuum
  deficit map and the ratio map of the observed to modeled radio continuum
  emission of NGC~4522 overlaid with radio spectral index contours
  taken from \citet{bv04}.  
  The contour levels increase by 0.15 and range between $0.7$ to
  $1.9$; the dashes indicate the direction of the downward
  gradient.  
  We plot the same images in the bottom panels except this time we
  overlay them with 6~cm polarized radio contours \citep{bv04}.
  The contours are logarithmically scaled and begin at
  the 3-$\sigma$ RMS level.
  The cross in each panel identifies the center of the galaxy.
  \label{fig-n4522}}
\end{figure}

\begin{figure}
\plotone{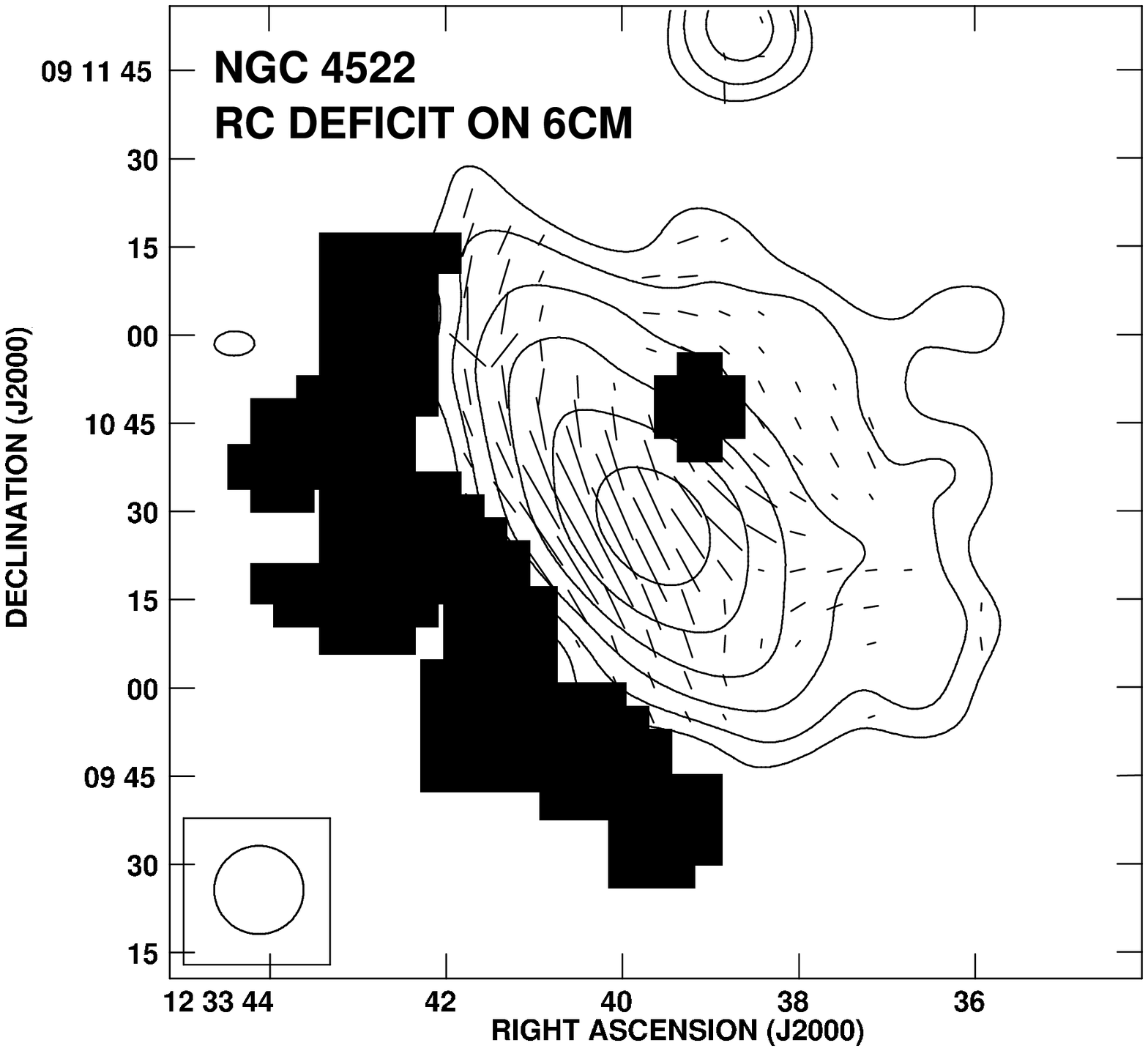}
\caption{The radio deficit map of NGC~4522 plotted with the 6~cm total continuum emissions as contours and the vectors of the magnetic field uncorrected for Faraday rotation \citep{bv04}.  
The contours are logarithmically scaled and begin at the 3-$\sigma$ RMS level.    
The size of the magnetic field vectors are proportional to the intensity of the polarized radio emission.  
The beam ($19\arcsec \times 19\arcsec$) is plotted in the bottom left corner of the image.  
\label{fig-n4522b}}
\end{figure}

Radio polarization and spectral index data of NGC~4522 provide evidence for an ongoing ICM-ISM interaction \citep{bv04};
the eastern edge of NGC~4522 is found to be highly polarized and has the flattest spectral index, which steepens as one moves towards the western side of the galaxy.  
In the first column of Figure \ref{fig-n4522} we plot the radio continuum deficit regions and overlay the 20 to 6~cm spectral index and 6~cm polarized radio continuum maps of \citet{bv04}.
We find that the regions of high polarization and flattest spectral index (i.e. $\alpha \sim 0.7$) lie just interior to the radio deficit region.    
While the peak in the polarized radio continuum is coincident with the flattest part of the spectral index distribution, this is not where the total radio intensity (or FIR) peaks as pointed out by \citet{bv04}.   
This is also evident in Figure \ref{fig-n4522b} where we show the 6~cm radio continuum emission together with the vectors of the magnetic field uncorrected for Faraday rotation \citep{bv04}, on the radio deficit map.  
The magnetic field vectors appear to be elongated parallel to the radio deficit region and the ICM-ISM interaction surface.    

These observations suggest that, except for synchrotron tails or the radio deficit regions, 
to first order, star formation in the disk still drives the appearance of the FIR/radio ratios.   
However, by inspecting the observed-to-modeled radio continuum ratio map of NGC~4522 displayed in the second column of Figure \ref{fig-n4522}, which was created using our {\it smoothed} 70~$\micron$ image, ICM-ISM effects become more apparent.    
We find a localized peak in the ratios just outside the radio deficit region, and near the regions of high radio polarization and flat spectral indices;
we attribute this peak to a local enhancement in the radio brightness.    
We now try to determine the most plausible physical scenario to explain this combination of observations.  

\subsection{Ram Pressure and the Relativistic ISM: Magnetic Field Compression, Shearing, and/or CR Re-acceleration? \label{sec-CR_B}}  
Ram pressure from the ICM can in principle work to both compress the
ISM, possibly triggering star formation \citep[e.g.][]{bd86,cald93},
as well as truncate a galaxy's star-forming disk by sweeping away
interstellar material \citep[e.g.][]{kk99}.
Similarly, ram pressure may act on a galaxy's relativistic ISM by
compressing or stretching its magnetic field, redistributing CR
particles, as well as by re-accelerating CR particles via ICM
triggered shocks.  

While the high level of radio polarization in NGC~4522 could arise
from a compressed magnetic field, such an occurrence is not clearly supported by our observations.  
Since the synchrotron emissivity goes roughly as the square of the magnetic field strength, we would expect a distinct region of enhanced radio continuum emission, where the field is being compressed and amplified, to lie just interior to the radio deficit region.
If there is such an enhancement in the total radio continuum just behind the radio deficit region we are probably not resolving it, but its amplitude must be modest since a strong signature in the ratio map is not detected.  


On the other hand, we also observe a large region of relatively flat spectral indices (i.e. $\alpha \sim 0.7$) interior to the deficit region. 
While \citet{nkw97} report an average radio spectral index of $0.83\pm0.02$ for a sample of 74 star-forming galaxies, the radio spectral indices of SNRs, whose associated shocks are thought to be the primary acceleration sites for CRs, are found to be slightly flatter, typically ranging between $\sim$$0.5-0.7$ both in the Galaxy \citep[e.g.][]{sr08,ahs08} and M~33 \citep[][]{nd95}.  
These values are also consistent with those predicted by diffusive shock acceleration theory \citep[e.g.][]{dmv89}.  
Thus, it seems plausible that the flat indices interior to the deficit region of NGC~4522 is the signature of CR electron acceleration sites within NGC~4522.  

We would therefore expect such areas to exhibit moderate enhancements in total continuum emission.  
These observations are consistent with the interpretation that the ICM wind is sweeping up CR electrons in a low pressure outer-region of the disk as well as driving a series of shocks into the ISM which will re-accelerate CR electrons there.      
The latter secondary shocks, or shocklets, will likely originate at the interface between the ICM and ISM, arising from inhomogeneities and turbulence.  
This would then explain the large region of flat spectral radio indices and area of modest increase to the radio continuum, as observed.  
Some of the CR electrons, including those that are newly accelerated, will be swept behind the galaxy creating a synchrotron tail that limit the enhancement to the radio surface brightness even though acceleration is occurring.   

A second explanation for the existence of the high radio
polarization could be that of shear as discussed by \citet{bv04} which
does not require total magnetic field amplification.
In this scenario the ICM wind sweeps up CRs and the magnetic field; 
the ISM magnetic field lines are stretched thereby increasing the fraction of polarized radio continuum emission without significantly increasing the total field strength.
This process alone, however, does not explain the observed flat
spectral indices.  

We therefore prefer the scenario in which the magnetic field of
NGC~4522 is being sheared along the interacting region between its
relativistic disk and the ICM as ram pressure induced shocks propagate
into its ISM.  
The locally depressed radio continuum emission in the deficit region
is the result of ram pressure from the ICM sweeping up low density CR
electrons, associated with past star formation activity which have
diffused significantly far away ($\sim$kpc) from their acceleration
sites in supernovae.  
While we do not find significant enhancements to the total radio continuum behind the deficit regions, the magnetic field may still be moderately compressed.  
Some of the CR electrons which have been swept-up by the ICM are redistributed downstream producing the synchrotron emission tails as observed very prominently here, as well as for NGC~4402 \citep{hc05}; this effect will limit the enhancement to the radio surface brightness.    
Another subset of particles, however, are re-accelerated interior to the radio continuum deficit region raising the total radio power as evidenced by the flat spectral indices and modest increase in the radio continuum found in the ratio map of NGC~4522.  
The preference for this scenario is discussed further in the context of the observations from the entire sample in $\S$6.2.  

\section{Discussion}
We have shown that a large fraction of our sample of cluster galaxies exhibits statistically significant radio deficit regions relative to the prediction of the phenomenological image-smearing model of \citet{ejm08}.   
Such deficit regions are not found within normal field galaxies, suggesting that cluster environment processes are at work.
Comparing our results with other observational data for NGC~4522, a galaxy which is clearly experiencing the effects of ram pressure, we find our deficit region agrees with such a scenario and 
adds new insight into the interaction.  

Similar to the example of NGC~4522, radio polarization observations of other Virgo cluster galaxies exhibit highly polarized radio emission along the side of the galaxy presumed to be the leading edge of the ICM-ISM interaction.  
\citep{bv07,kc07}.  
For the galaxies in our sample having polarization data and which show evidence for being disturbed (i.e. NGC~4254, NGC~4388 and NGC~4402), the regions of highly polarized continuum emission appear coincident with the modest local radio continuum enhancements found in our ratio maps (i.e. ratios of $\sim$$1.1-1.7$) and interior to our radio deficit regions.  
When averaged over these kpc-sized regions, such enhancements in the total radio continuum translate, roughly, into a compression and amplification of the magnetic field by $\sim$$5-30$\%, clearly a modest effect.  
Thus, the scenario described for NGC~4522 seems to be applicable for
these galaxies as well based on the polarized radio continuum results.   
With this picture we will use the findings from our analysis to
quantitatively assess the strength of the ram pressure from the
intracluster medium (ICM). 
Since the radio deficit regions detected for NGC~4396 and NGC~4579 differ empirically from what we find for the rest of the sample, and show evidence for being artificial, we exclude them from the following discussion (see $\S$\ref{sec-ratmaps}).

\begin{deluxetable*}{cccccccccc}
  \tablecaption{Radio Continuum Deficit Regions Results
  \label{tbl-rcdefres}}
\tablecolumns{10}
\tablewidth{0pt}
\tabletypesize{\scriptsize}
  \tablehead{
    \colhead{} & 
    \colhead{$S_{\nu}^{\rm obs}$} & \colhead{$S_{\nu}^{\rm mod}$} & 
    \colhead{} & \colhead{$A_{\rm def}$} & \colhead{} &
    \colhead{$P_{\rm RISM}^{\rm mod}$} & \colhead{$P_{\rm RISM}^{\rm
    obsdisk}$} & \colhead{} & \colhead{$t_{\rm quench}^{a}$} \\
    \colhead{Galaxy} & \colhead{(mJy)} & \colhead{(mJy)} &
    \colhead{$\Upsilon$} & \colhead{(kpc$^2$)} & \colhead{$A_{\rm
	def}/A_{\rm disk}$} & \colhead{(10$^{-12}$~dyn~cm$^{-2}$)} &
    \colhead{(10$^{-12}$~dyn~cm$^{-2}$)} & 
    \colhead{$P_{\rm RISM}^{\rm mod}$/$P_{\rm RISM}^{\rm obsdisk}$} &
	\colhead{(Myr)}}
  \startdata
  \cutinhead{Virgo (Cluster) Sample}
  n4254&    1.578&    4.728&  0.007&  33.0&  0.04&     3.9&     8.3&    0.46&\nodata\\
  n4330&    1.891&    4.939&  0.153&  20.7&  0.27&     2.0&     2.1&    0.97&\nodata\\
  n4388&    1.733&    7.607&  0.036&  22.6&  0.21&     2.1&     5.6&    0.38&  225\\
  n4402&    1.120&    4.134&  0.045&  18.8&  0.13&     2.3&     4.3&    0.53&  200\\
  n4522&    1.676&    4.228&  0.096&  18.3&  0.17&     2.5&     3.2&    0.80&  100\\
  n4569&    1.852&    4.702&  0.023&  35.6&  0.12&     2.6&     7.1&    0.36&  300\\
\cutinhead{Field (Control) Sample}
  NGC~3627&    2.207&    7.318&  0.011&  43.3&  0.08&     2.9&     6.8&    0.43 &\nodata
  \enddata
\tablecomments{NGC~4321 and NGC~4580, along with nearly every field galaxy, are excluded since radio continuum deficit regions were not found within their disks; we note that the $P_{\rm RISM}^{\rm obsdisk}$ values for these two galaxies are 7.1 and $16.0\times 10^{-12}$~dyn~cm$^{-2}$, respectively.  NGC~4396 and NGC~4579 were excluded since their radio deficit regions may be artificial (see $\S$
\ref{sec-ratmaps}).\\
$^{a}$: Time since outer-disk star formation was quenched taken from \citet{hc08}; for NGC~4580 they report $t_{\rm quench} = 475$~Myr.  
} 
\end{deluxetable*}

\subsection{Estimates of Minimum Ram Pressure and Internal
  Relativistic ISM Pressure \label{sec-pram}} 
Ram pressure is expected to vary by several orders of magnitude as galaxies orbit within clusters \citep{bv01,tbvg07}, but there is no known way to measure the strength of current ram pressure on a galaxy.
The radio deficit offers the potential to do this.

\begin{figure}
\plotone{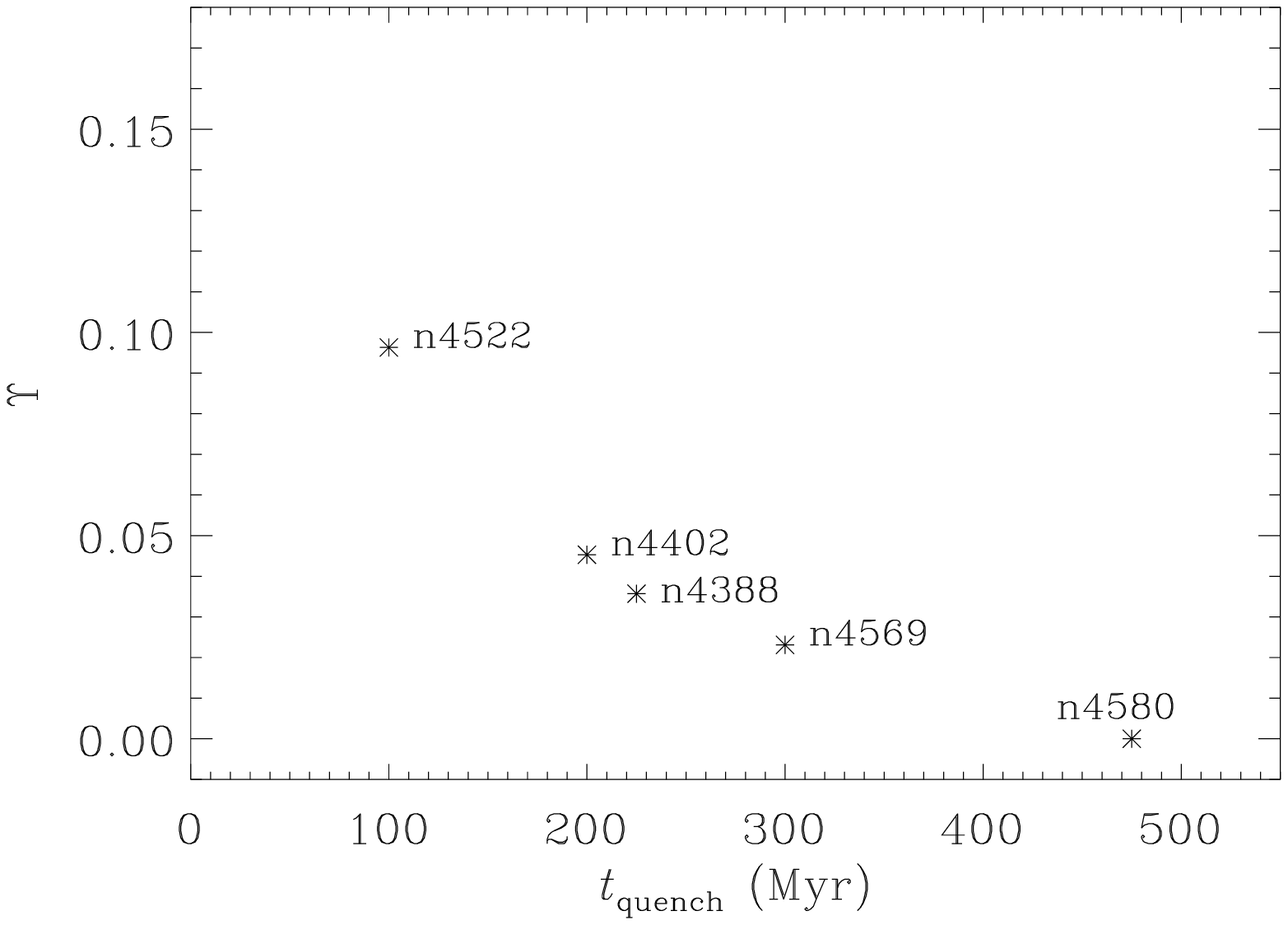}
\caption{The severity of the deficit region characterized by
  $\Upsilon$ (see $\S$\ref{sec-rcdef-def}) plotted against time since star formation was quenched in their outer disks \citep{hc08}, likely coinciding with the time since the galaxy disk experienced peak-pressure.  \label{fig-tUp}}
\end{figure}

Figure \ref{fig-tUp} shows that the strength of the radio deficit is closely related to the time since peak pressure as inferred from stellar population studies and gas stripping simulations \citep{hc08, bv03, bv04b, bv06}. 
This strongly suggests that the strength of the radio deficit is good indicator of the strength of the current ram pressure.
In simple models \citep[e.g.][]{bv01} the ram pressure drops
by roughly an order of magnitude 300 Myr after peak pressure, and by a factor of $\sim$5 from 100 Myr to 300 Myr after peak pressure.
As shown in Figure \ref{fig-tUp}, this is similar to the change in the radio deficit parameter between NGC~4522 and NGC~4569, suggesting that differences from galaxy to galaxy in the  radio deficit parameter are reflecting differences in the strength of current ram pressure.

We can get a rough estimate of the strength of ram pressure acting on these galaxies from the equipartion energy density of the radio-emitting plasma.
Ram pressure from the ICM is simply defined as, 
\begin{equation}
\label{eq-pram}
P_{\rm ram} = \rho_{\rm ICM} v_{\rm gal}^{2}
\end{equation}
where $\rho_{\rm ICM}$ is the ICM mass density and $v_{\rm gal}$ is
velocity of a galaxy relative to the ICM.
If the ICM ram pressure exceeds the internal pressure of a galaxy's 
relativistic ISM (CRs $+$ magnetic field), then it should be possible to redistribute
and even strip them from the galaxy disk.  
Using the predicted radio flux density for each deficit region, we can
approximate a minimum value for the ICM ram pressure needed to cause the
observed depression in the radio.  

Taking the predicted flux density of the deficit region along with its
area we use the revised equipartition and minimum energy formulas of
\citet{bk05} to calculate the minimum energy magnetic field strength
of the deficit regions before they were stripped. 
This calculation assumes a proton-to-electron number density ratio, for particles in the energy range corresponding to GHz synchrotron emission, of 100, a radio spectral index of $0.8$, and a path length through the emitting medium of $1/cos(i)$~kpc where $i$ is the galaxy
inclination. 
The inclination is derived using the method of \citet{dd97} such that, 
\begin{equation}
\cos^{2}i = \frac{(b/a)^{2} - (b/a)^{2}_{\rm int}}{1-(b/a)^{2}_{\rm
    int}},
\end{equation}
where $a$ and $b$ are the observed semi-major and semi-minor axes and
the disks are oblate spheroids with an intrinsic axial
ratio $(b/a)_{\rm int} \simeq 0.2$ for morphological types earlier
than Sbc and $(b/a)_{\rm int} \simeq 0.13$ otherwise.
Using these calculated magnetic field strengths $B$ (i.e. calculated using
Equation 4 of \citet{bk05}) we compute the magnetic field energy
densities $U_{\rm B} = B^2/(8\pi)$ of the deficit regions.  

Assuming minimum energy between the magnetic field and CR particle
energy densities, $U_{\rm B}$ and $U_{\rm CR}$ respectively, we can
use the values of $U_{\rm B}$ over the deficit region in our model to
determine the minimum $P_{\rm ram}$ necessary to create the deficit
regions.    
From minimum energy arguments we find that $U_{\rm B} = 3/4U_{\rm
  CR}$. 
For a relativistic gas the magnetic pressure, $P_{\rm B}$, and CR
pressure $P_{\rm CR}$ are related to energy density such that $P_{\rm
  B} = U_{\rm B}$ and $P_{\rm CR} = 1/3U_{\rm CR}$ leading to the
relation that $P_{\rm CR} = 4/9 U_{\rm B}$.
Then, the pressure of the relativistic ISM is found to be 
\begin{equation}
\label{eq-PrISM}
P_{\rm RISM} = P_{\rm CR} + P_{\rm B} \sim 13/9 U_{\rm B}.
\end{equation}
The relativistic ISM pressure of the deficit regions before they were stripped, as estimated from our models and denoted as $P_{\rm RISM}^{\rm mod}$, are given in Table
\ref{tbl-rcdefres};
these estimates set the minimum ICM ram pressure necessary to create the
observed radio continuum deficit regions (i.e. $P_{\rm ram} > P_{\rm
  RISM}^{\rm mod}$).  

\begin{figure}
\plotone{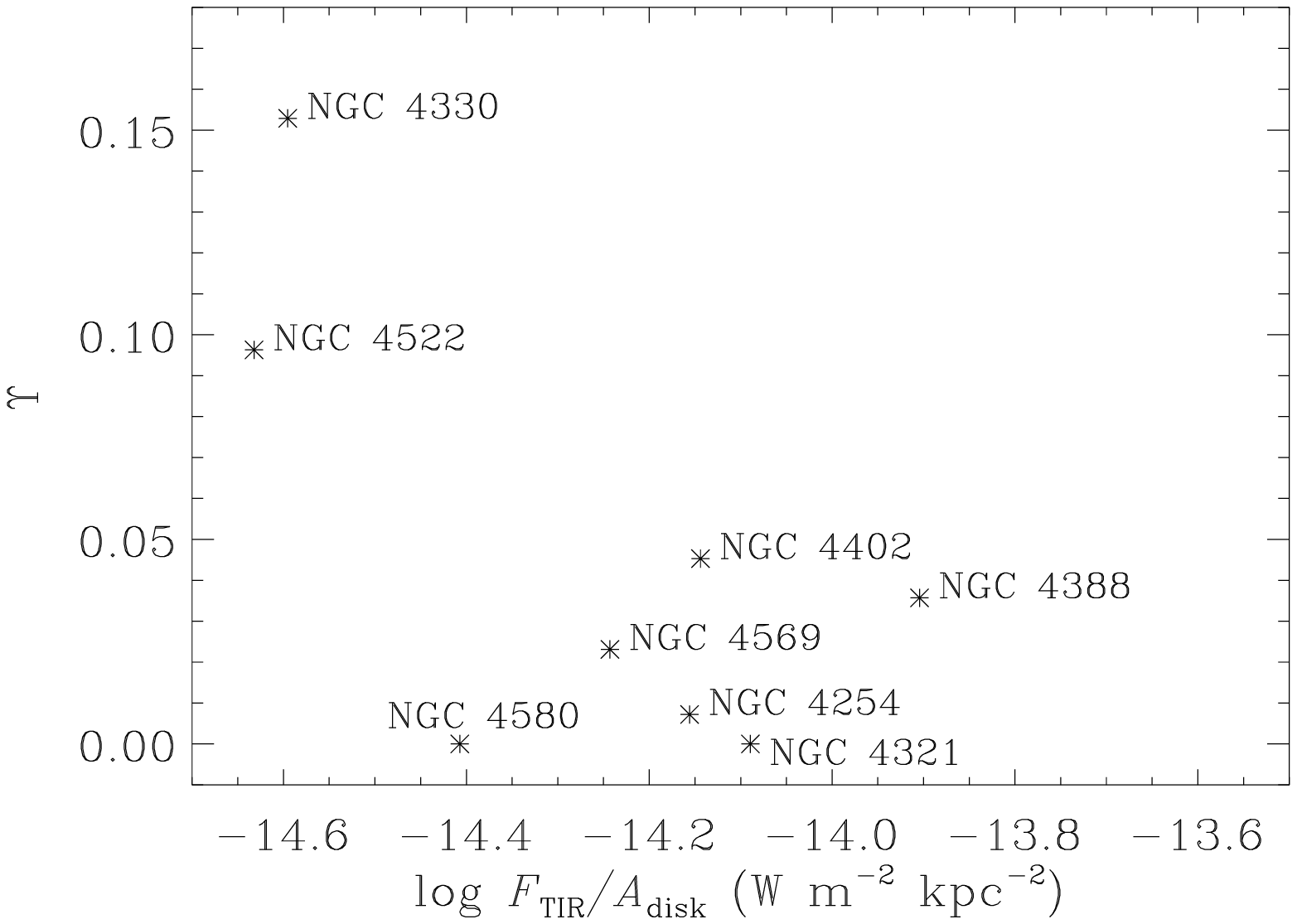}
\caption{The severity of the deficit region characterized by $\Upsilon$ (see $\S$\ref{sec-rcdef-def}) plotted against a measure of TIR surface brightness. 
\label{fig-TIRUp}}
\end{figure}

All of the Virgo galaxies listed in Table \ref{tbl-rcdefres} have $P_{\rm RISM}^{\rm mod}$ values ranging between $\sim$$2 - 4 \times 10^{-12}$~dyn~cm$^{-2}$.
Using the projected linear distances to the cluster center for these sample galaxies along with a measured density profile of Virgo \citep{mats00} yields a range in ICM density of  $n_{\rm ICM} = \rho_{\rm ICM}/m_{p} \approx 1.3- 4.2\times10^{-4}$~cm$^{-3}$.  
Taking a Virgo galaxy velocity of 1000~km~s$^{-1}$, which is comparable to the $\sim$930~km~s$^{-1}$ velocity dispersion among Virgo spirals \citep{bts87} and typically achieved by cluster members during some portion of their orbits \citep{bv01}, the associated range in ICM ram pressure is $P_{\rm ram} \approx 2 - 7\times 10^{-12}$~dyn~cm$^{-2}$.  
This range of ICM pressures is generally greater than or similar to the values for $P_{\rm RISM}^{\rm mod}$ which is consistent with ICM ram pressure being able to create the observed deficit regions.
These pressures agree with those produced in the 3D hydrodynamical simulation of \citet{ermb07} over the same projected linear distances to the cluster center among our sample.
We also note that, given the ICM density range, the corresponding velocities required to reach a minimum ram pressure of $\sim$$2\times10^{-12}$~dyn~cm$^{-2}$ range between 530 and 980~km~s$^{-1}$, quite comparable to the velocity dispersion of Virgo spirals.  

While this value only sets the minimum ram pressure strength required
to create the deficit region, it is perhaps more interesting to
try to get a better handle on the equilibrium pressure at the working
surface between the deficit region and the edge of the galaxy's
observed radio disk.  
Since it is difficult, and probably not very accurate, to calculate the
relativistic ISM pressure on a point-by-point basis within the disk, we
instead measure the average internal relativistic plasma pressure of
observed radio disk $P_{\rm RISM}^{\rm obsdisk}$.
We define $P_{\rm RISM}^{\rm obsdisk}$ using the observed total radio
flux density and the area given within the 3-$\sigma$ RMS isophote of
each radio image.
The values of $P_{\rm RISM}^{\rm obsdisk}$, along with the ratios
between $P_{\rm RISM}^{\rm mod}$ and $P_{\rm RISM}^{\rm obsdisk}$, are
given in Table \ref{tbl-rcdefres}.
While NGC~4321 and NGC~4580 were excluded in Table \ref{tbl-rcdefres}
since they do not have radio deficit regions, we note that their
values of $P_{\rm RISM}^{\rm obsdisk}$ are 7.1 and $16.0\times
10^{-12}$~dyn~cm$^{-2}$, respectively.
Therefore, the full range of relativistic plasma pressures are
similar to ram pressures calculated above for a typical Virgo galaxy
and a range of ICM densities.

The internal relativistic plasma pressures cannot be vastly different
from that of the ICM ram pressure since then we would either not
observe the radio deficit regions or the entire relativistic ISM would
be swept out of the disk.  
We expect to see more significant radio deficit regions  (i.e. larger
values of $\Upsilon$) for cases where the internal relativistic ISM
pressure is low.
Comparing the values of $\Upsilon$ with $P_{\rm RISM}^{\rm obsdisk}$
in Table \ref{tbl-rcdefres} we find that this is indeed the case.
Our ram pressure estimates from minimum energy arguments give a good rough approximation for the strength of ram pressure within a factor of a few, but cannot be expected to measure precisely the differences in ram pressure among the sample galaxies.
Further work is required to more closely and quantitatively relate the radio deficit to the strength of ram pressure.  

In Figure \ref{fig-TIRUp} we find that there is a general trend of decreasing values of $\Upsilon$ with increasing TIR surface brightnesses as measured by the ratio of each galaxy's global TIR flux to $A_{\rm disk}$, the deprojected area of their ratio map.    
Since TIR surface brightness and $P_{\rm RISM}^{\rm obsdisk}$ are directly related, this trend is not terribly surprising.  
Interpreting the TIR surface brightness to be a proxy for the mean ISM density, 
we find that galaxies having a more dense ISM appear to be less affected by the ICM.
This seems reasonable given that the ram pressure would need to be stronger to disturb their ISM and is consistent with higher values for the internal relativistic plasma pressure for small values of $\Upsilon$.    
However, this trend may be fortuitous given that the external ram pressure likely varies tremendously among the sample.  



\subsection{Combining Local and Global Observations into a Single Physical Scenario}
Using our results on the spatially resolved FIR/radio distributions of these cluster spirals, along with their global FIR/radio properties,  we will now investigate whether the ICM wind plays a significant role in enhancing the escape and re-acceleration of CR electrons.   
After considering escape, we explore a number of physical scenarios to explain simultaneously the local and global FIR/radio ratio behavior among these galaxies before coming to our preferred picture.  

\subsubsection{Cosmic Ray Electron Escape}
\begin{figure}
\plotone{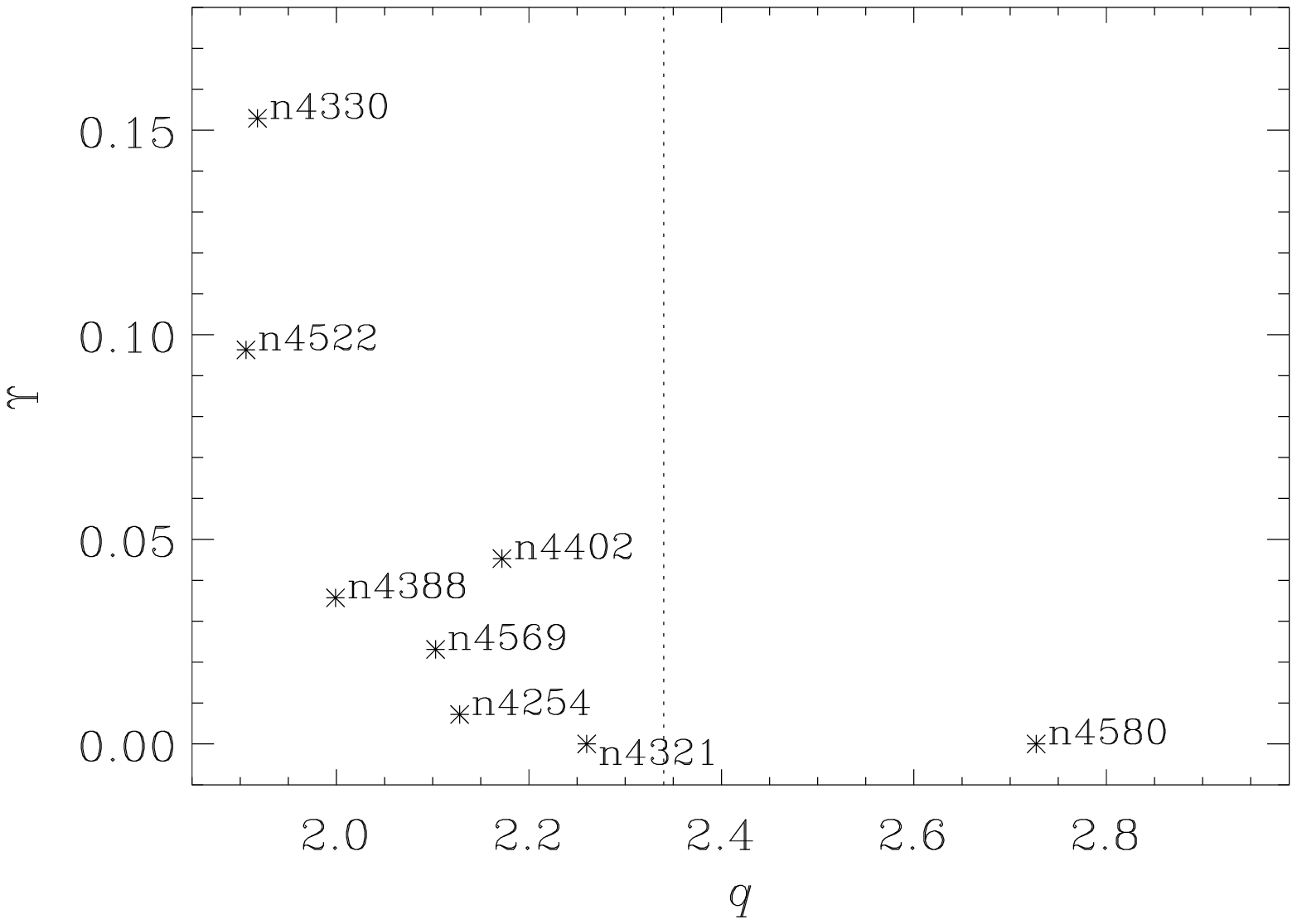}
\caption{The severity of the deficit region characterized by
  $\Upsilon$ (see $\S$\ref{sec-rcdef-def}) plotted against $q$ (the
  logarithmic FIR/radio ratio). The vertical line at $q = 2.34$ identifies the average $q$ value reported by \citet{yrc01} for 1809 galaxies.  
  \label{fig-qUp}}
\end{figure}

The only galaxy with a high global $q$ value is NGC~4580.  
While the CR electrons are being moved around by the ICM wind, it is unclear as to whether 
they easily escape the galaxy disks as global $q$ (logarithmic FIR/radio) ratios do not appear systematically high with respect to the nominal value of $\sim$$2.34 \pm 0.26$~dex \citep[i.e.][]{yrc01}.  
In fact, by plotting $q$ against $\Upsilon$ (our parameter defining the severity of the deficit region) in Figure \ref{fig-qUp} we find a trend of decreasing FIR/radio ratios with increasing $\Upsilon$ and that nearly all of the $q$ values are lower than the nominal value.   
Other, more detailed studies on the global FIR and radio properties of cluster galaxies have pointed out notably lower FIR/radio ratios than expected from the FIR-radio relation in the field \citep[][]{mo01,ry04}.  
This suggests that there is either depressed FIR emission or an enhancement in the global radio emission among our sample galaxies.  
And while the trend in Figure \ref{fig-qUp} may simply be the result of small number statistics, it also might suggest that $q$ is sensitive to the strength of the ICM ram pressure and perhaps a galaxy's stripping history.

We do not detect a radio deficit region for NGC~4580. 
This galaxy is highly H~{\sc i} deficient, having the largest deficiency value in Table \ref{tbl-galdat} (1.53~dex; a factor of  $\sim$5 larger than the median value among the sample), 
and has been classified as an anemic spiral \citep{svb76}, though better described as having a severely truncated star-forming disk \citep{kk04}.   
It appears to have been stripped long ago ($\sim$475~Myr) and is the only post-strongly stripped galaxy in our sample observed so long after peak pressure \citep{hc08}. 
We therefore speculate that its high $q$ value is the result of CR electron escape as the ICM wind has swept out most of the galaxy's gaseous and relativistic ISM long ago leaving only the central disk.  
The only other galaxy for which we do not detect a radio deficit region is NGC~4321 which is much different from NGC~4580.
NGC~4321 has not been strongly stripped and has a near-average $q$ value; this galaxy appears much more similar to the WSRT-SINGS field galaxies.  

\subsubsection{Low Global FIR/Radio ratios as a Result of Re-acceleration?} 
As already discussed in $\S$\ref{sec-CR_B}, we have attributed local enhancements to the radio continuum arising from the re-acceleration of CR electrons by ICM induced shocklets.  
While moderate local enhancements are found in the radio continuum interior to the radio deficit regions via the ratio maps, they do not appear significant enough to explain $q$ ratios which are $\la$0.4~dex lower than average (i.e. a factor of $\la$2.5).  
The $q$ values of 'Taffy' systems are found to be a factor of $\sim$2
low due to excess radio continuum emission associated with the
synchrotron bridges connecting the galaxy pairs  \citep{chss93,chj02}.     
Looking to see if synchrotron tails may be causing a similar effect in
NGC~4522 (where the observed synchrotron tail appears most striking)
we find that this cannot be the case.  
Only $15 - 20$\% of the total radio continuum emission is associated
with the tail.  

While we have argued that it does not seem likely that the magnetic field has been compressed significantly (i.e.  no more than $\la$30\% as suggested by the radio excesses) at the leading edge between the ISM and the ICM wind, leaving no clear signatures of a gradient in the ratio map, perhaps the field strength is raised globally due to thermal compression by the surrounding ICM as suggested by \citet{mo01}.  
Models of such an effect seem to be able to produce the factor of $2-3$ excess synchrotron emission and reproduce the observed $q$ ratios \citep{ry04}.  
If such a scenario were true, and the Virgo galaxies with low $q$ ratios are simply compressed versions of field galaxies, we would expect galaxies with the lowest $q$ ratios to appear compressed in all directions and exhibit relatively larger internal relativistic plasma pressures, $P_{\rm RISM}^{\rm obsdisk}$, compared to Virgo galaxies with larger $q$ ratios.      
However, the galaxies having the lowest $q$ ratios both exhibit extraplanar radio continuum emission along their trailing edges in the form of diffuse synchrotron tails and do not have larger $P_{\rm RISM}^{\rm obsdisk}$ values relative to the rest of sample, inconsistent if compression was the dominant mechanism.    

Another possibility is that the FIR emission is depressed in these systems through the removal of a dust component.
For instance, the ICM wind may have removed a substantial amount of the cool dust (cirrus) component, a significant contributor to the total FIR emission.
However, FIR color differences between the stripped Virgo galaxies and the WSRT-SINGS spirals are too small to induce differences in the SEDs and total FIR emission large enough to account for the differences in $q$.  
The MIPS  $f_{\nu}(70~\micron)/f_{\nu}(160~\micron)$ and $f_{\nu}(24~\micron)/f_{\nu}(70~\micron)$ flux density ratios have medians that are $<$~20\% lower and higher, respectively, than those found for the WSRT-SINGS spirals.  
Both samples have FIR colors in the range of quiescent normal star-forming galaxies, where the 20\% color differences translate into small changes in the SED, and changes of only a few percent in FIR normalized either to the 24 or 70~$\micron$ emission, based on the \citet{dd02} SED templates.  
The 24~$\micron$ emission component is relevant for normalization because of its close association with HII regions and dense, star-forming regions \citep{dc07}, making it  much more resistant to stripping than the longer wavelength components.
This result is consistent with the study of \citet{ry04} who also found that the FIR properties (i.e. luminosities and colors) of cluster galaxies having excess radio emission are similar to those of field galaxies.

The FIR emission in these systems may also be depressed, without systematic changes in their FIR colors, as a result of depressed star formation activity per unit area.
If ram pressure from the ICM wind is driving shocklets into the ISM, mechanical energy would be dissipated into molecular clouds as turbulence, possibly reducing their efficiency to collapse and form new stars.
However, in order for a galaxy to exhibit $q$ values similar to those reported here, the decrease in star formation activity must  have begun within the last $\sim$50~Myr (i.e. roughly the expected CR electron cooling time to synchrotron losses).   
At later epochs, the synchrotron emission associated with the previous star formation episodes fades away, and $q$ values drift back towards more normal values even if the star formation activity remains depressed.    
Since 50~Myr is more than an order of magnitude smaller than the cluster crossing time, it is unlikely that we would be observing so many galaxies during this special period of time.
It therefore appears that depressed FIR emission is not  the dominant effect driving the systematically low $q$ values for these galaxies.

While it is still not completely unambigous, we believe the most plausible explanation is that the ICM wind raises the global CR electron energy and synchrotron power by driving shocklets into the ISM
which are re-accelerating CR electrons.  
Taking again $v_{\rm gal} \approx 1000$~km~s$^{-1}$ and the range in $\rho_{\rm ICM}$ values discussed in $\S$ \ref{sec-pram}, and assuming the average surface area of each galaxy disk exposed to the ICM wind is 25~kpc$^{2}$ (i.e. the mean size of the radio deficit regions given in Table \ref{tbl-rcdefres}), we compute the kinetic energy per unit time a galaxy will inherit from the ICM wind.  
Comparing this number to that of the mean radio continuum luminosity among those galaxies listed in Table \ref{tbl-rcdefres}, and assuming a number density ratio of protons-to-electrons, for particles in the energy range corresponding to GHz synchrotron emission, of 100, we find that $\sim$$3 -11$\% of the total kinetic energy available must be given to the CRs in order to double the radio luminosity.    
While a value $\sim$11\% may be slightly large, there are many uncertainties involved with this calculation including whether the CR electrons are more efficient than CR nuclei at absorbing this energy.   
In any case, it is certainly clear that the available mechanical luminosity of the system is enough to cause the observed global excess in synchrotron emission.  

The fact that we only find relatively moderate local enhancements to the total continuum behind the radio deficit region, save perhaps NGC~4254 where a less modest radio excess is observed (i.e. ratios peaking at $\sim$1.7), suggests that the shocks run through the entire galaxy disk rather quickly.  
The shock speed through the galaxy, $v_{\rm s}$, which must be super-Alfv\'{e}nic to re-accelerate CR electrons, should have a value around 
\begin{equation}
\label{eq-vshock}
v_{\rm s} \approx \left(\frac{4}{3}\frac{P_{\rm ram}}{\rho_{\rm ISM}}\right)^{1/2} \approx
v_{\rm gal} \left(\frac{4}{3}\frac{\rho_{\rm ICM}}{\rho_{\rm ISM}}\right)^{1/2}, 
\end{equation}
where $P_{\rm ram}$ is the ram pressure, as defined in Equation \ref{eq-pram}, and $\rho_{\rm ISM}$ and $\rho_{\rm ICM}$ are the ISM and ICM densities, respectively.
Again taking $v_{\rm gal} \approx 1000$~km~s$^{-1}$, the range in $\rho_{\rm ICM}$ values discussed in $\S$ \ref{sec-pram}, and $n_{\rm ISM} = \rho_{\rm ISM}/m_{p} \approx 1$~cm$^{-3}$  leads to shock velocities ranging between $\sim$15 to 25~km/s.  
Assuming a thin disk thickness of 500~pc, the shocks should run through each disk within $\sim$20 to 40~Myr; indeed this is very short compared to the dynamical timescale of these systems.  

As previously noted, this proposed mechanism may work to slow down star formation itself, decreasing the infrared surface brightness,  either from an increase in cold molecular cloud heating by the re-accelerated CR particles or similar gas-phase shocklets perturbing the clouds.  
However, it is unclear how significant this effect may be relative to the increase in synchrotron emissivity via the associated re-acceleration of CR electrons.   

\begin{figure}
\plotone{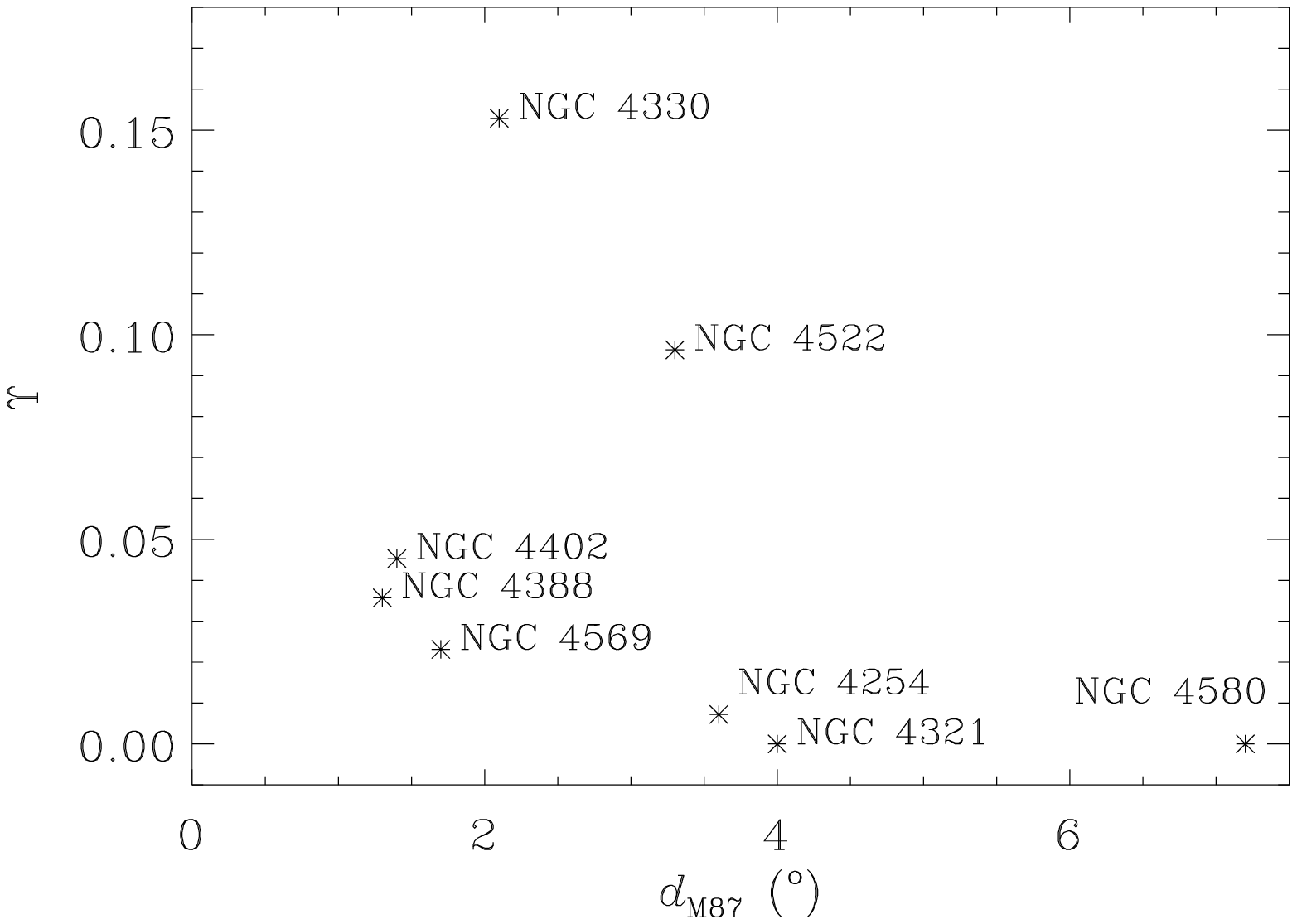}
\caption{The severity of the deficit region characterized by
  $\Upsilon$ (see $\S$\ref{sec-rcdef-def}) plotted against the
  distance, in degrees, to the Virgo cluster center.\label{fig-dUp}}
\end{figure}

\subsection{Pre- or Past-Peak Pressure? \label{sec-peakp}}
Since many cluster galaxies have highly radial orbits, the direction
to the cluster center can help provide some insight as to whether a
galaxy may be traveling into or away from the cluster center.
The direction toward the cluster center 
is indicated on the radio deficit maps in Figure \ref{fig-maps}.
Some galaxies have deficit regions facing toward the cluster center
(NGC~4254, and NGC~4402) which is suggestive of a galaxy entering the cluster and at a phase of pre-peak pressure.
On the other hand, we also find an example for which the radio deficit
region faces away from the cluster center (NGC~4569) which may suggest
that the galaxy is on their way out of the cluster and past
peak-pressure.    
There are also galaxies which face neither directly towards or away
from the cluster center (NGC~4330 and NGC~4522) leading to a
more complex scenario;
such objects may be stripped by interactions with sub-cluster
structures (e.g. NGC~4522  with M~49).
We also note that NGC~4388 may be interacting with the M~86
sub-cluster;
its H~{\sc i} tail points neither towards or away from the cluster
center but rather towards M~86 \citep{ovg05}.

Another interesting case is that of NGC~4254 which is thought to be
entering Virgo for the first time at high speed.
It has recently been suggested by \citet{bk07} and \citet{hgk07} that
a large H~{\sc i} tail associated with NGC~4254 is a tidal arm
resulting from galaxy harassment \citep{bm96, mlk98}.  
However, no interacting counterpart has been identified.
Also, the H~{\sc i} tail is located opposite the observed radio
deficit region and the direction to the cluster center.
Since it appears that the radio deficit regions are related to the
effects of ram pressure, it therefore seems plausible that the large
H~{\sc i} tail  could be the result of ram pressure stripping.   
However, it is also possible that the radio deficit region may be
created via galaxy harassment or that, more plausibly, both physical processes
are currently at work.  

Projection effects, specifically the orientation of the galaxy disk with respect to the ICM wind direction and our line-of-sight, likely play a significant role in determining the location of deficit region centroids. 
For largely face-on systems, the relationship between the deficit region location and the direction of the wind is probably more easily predicted than for more highly inclined galaxies.  
The highly inclined galaxies NGC~4330, NGC~4402, and NGC~4522, have radio deficit regions consistent with lunes along one side of their disks;
these could occur over a large range of wind angles.
Furthermore, the size of the observed deficit region is subject to projection effects as well, likely appearing much smaller for disks viewed face-on, especially if they are moving through the ICM along the line-of-sight.  

Therefore it is hard in general to gauge whether galaxies may be pre- or past-peak pressure from the direction of the cluster center with respect to the deficit region.   
In Figure \ref{fig-dUp} we plot $\Upsilon$ against the distance to the cluster center.
The lack of observed trend suggests that the 3D structure of the cluster along with presence of sub-cluster structures are important in understanding ICM-ISM interactions and the picture of a smoothly distributed and static ICM is overly simplistic for characterizing ICM-ISM interactions.  

\subsection{Departures from Typical ISM Conditions}
While we can explain regions having radio continuum emission without corresponding FIR emission as the result of CR electrons being swept into a synchrotron tail by the ICM, the observation of FIR emission
without corresponding radio emission is a bit more interesting since it suggests a strong departure from typical ISM conditions.  
Even though such regions are close to the noise limits of the infrared and radio maps, this result appears to be significant.  
In Figure \ref{fig-pixid} we show two examples (i.e. NGC~4402 and NGC~4522) for which we detect 70~$\micron$ emission more radially extended than the radio; these regions are found {\it only} along the galaxy edge associated with the radio deficit region.
Furthermore, if $q$ were constant across these galaxy disks, we would expect to detect the radio continuum in these regions at $\ga$3-$\sigma$ RMS level of the radio maps.   
However, we know that $q$ does not remain constant across galaxy disks and actually decreases with increasing galactocentric radius \citet{ejm06a}, implying even higher levels of radio continuum should be expected, easily exceeding $\ga$6-$\sigma$ RMS noise levels.    
We therefore conclude that the non-detection of radio emission is significant at least at the $\sim$6-$\sigma$ level or, in other words, $\la$15\% of the expected radio continuum emission is present.  

These observations then suggest that the diffuse relativistic and gaseous ISM have been stripped away along these leading edges leaving the denser molecular material as traced by the 70~$\micron$ dust emission.  
Presumably, the low density relativistic phase is quickly removed due to an increase in open magnetic field lines leading to the quick escape of CR particles.  
While our observations do not rule out that synchrotron emission associated with the remaining molecular clouds is still present and simply below our detection limit, since the diffuse component will dominate the total radio continuum emission in these regions, it is still interesting to speculate how quickly this synchrotron emission associated with the remaining molecular clouds would fade as CR electrons diffuse away.    
%

Assuming a flux freezing scaling of $B \propto n^{1/2}$ \citep[e.g.][]{ruz88, nb97, cr99}, a solar neighborhood ISM density and magnetic field strength of $n_{\rm SN} = \rho_{\rm SN}/m_{p} \approx 1$~cm$^{-3}$ and $B_{\rm SN} \approx 6~\mu$G \citep{smr00}, and an average giant molecular cloud (GMC) density of $n_{\rm GMC} = \rho_{\rm GMC}/m_{p} \approx 10^{3}$~cm$^{-3}$ leads to a GMC magnetic field strength of $B_{\rm GMC} \approx 190~\mu$G.
This value is actually quite consistent with field strengths of molecular clouds inferred from the Chandrasekhar-Fermi technique \citep{cf53} using polarized dust continuum emission \citep[e.g][]{cr04}. 
Rewriting Equation 24 from \citet{ejm08} for diffusion alone, such that 
\begin{equation}
\label{eq-tdiffl}
\left(\frac{\tau_{\rm diff}}{\rm yr}\right) = 2 \times 10^{6} \left(\frac{l_{\rm diff}}{\rm kpc}\right)^{2} 
							    \left(\frac{\nu}{\rm GHz}\right)^{-1/4}
							    \left(\frac{B}{\mu{\rm G}}\right)^{1/4},
\end{equation}
where $\tau_{\rm diff}$ is the time it takes a CR electron emitting at frequency $\nu$ to diffuse a distance $l_{\rm diff}$ in a magnetic field of strength $B$ assuming random walk diffusion characterized by an energy-dependent diffusion coefficient, we can calculate the time it would take the CR electrons to diffuse away from the molecular clouds.
Taking $B = B_{\rm GMC} \approx 190~\mu$G and $l_{\rm diff} \approx 0.05$~kpc leads to a diffusion time of $\sim$1.7$\times 10^{4}$~yr for CR electrons to leave the molecular clouds; this is considerably short compared to the stripping timescales and plausible given our observations.    

Such a scenario may have implications for star formation in these regions as well.  
According to Equation \ref{eq-PrISM} and assuming minimum energy arguments, the complete lack of CR pressure, which works to support a cloud from collapse, will decrease the total relativistic ISM pressure by $\sim$30\%.
This situation may encourage cloud collapse in these regions leading to enhanced star formation.  
On the other hand, the ICM wind will also perturb molecular clouds introducing additional turbulence further discouraging collapse.
Determining which of these competing effects will actually dominate is beyond the scope of this paper.  

\section{Conclusions}
We have studied the interstellar medium (ISM) of 10 Virgo galaxies included in the VLA Imaging of Virgo in Atomic Gas (VIVA) survey using {\it Spitzer} MIPS and VLA 20~cm imagery.  
By comparing the observed radio continuum images with modeled distributions (i.e. appropriately smoothed FIR images), created using a phenomenological image-smearing model described by \citet{ejm06a, ejm08}, we find that the edges of many cluster galaxy disks are significantly radio deficient. 
These radio deficit regions are consistent with being leading-edge regions affected by intracluster medium (ICM) induced ram pressure as suggested by the location of H~{\sc i} and radio continuum tails.
From our results we are able to conclude the following:
\begin{enumerate}
\item
  The distributions of radio/FIR ratios within cluster galaxies thought to be experiencing ICM-ISM effects are systematically different from the distributions in field galaxies.  
  Radio/FIR ratios are found to be systematically low, due to a deficit of radio emission, along the side of the galaxy experiencing the ICM wind.  
  
\item
  In the case of NGC~4522, a clear example of ongoing strong stripping, we find that the radio deficit region lies exterior to a region of high radio polarization and a flat radio spectral index.   
  We interpret this to suggest that CR electrons in the halos of galaxies are being swept up by the ICM wind.   
  The ICM wind drives shocklets into the ISM of the galaxy which re-accelerate CR particles interior to the working surface at the ICM-ISM interface.
  Some CR electrons may also be redistributed downstream creating synchrotron tails as observed for NGC~4522. 
  The high radio polarization is probably mostly the result of shear as the ICM wind stretches the magnetic field;
  compression may also play a role, though a modest one, since the total radio continuum in these regions does not appear significantly enhanced.

 \item
The severity of the current ICM-ISM interaction, as measured by the normalized difference between the modeled and observed radio flux densities in the radio deficit regions, appears to be strongly correlated with the time since peak pressure as measured by stellar population studies and gas stripping simulations.
Thus the radio deficit appears to be a good tracer for the strength of current ram pressure.    


\item
  Using the identified radio deficit regions we are able to get a quantitative estimate of the minimum strength of the ICM pressure required to affect a galaxy disk; we find values in the range of $\sim$$2-4\times 10^{-12}$~dyn~cm$^{-2}$.  
  These pressures are generally smaller than, but similar to, those estimated for typical values of the ICM gas density and galaxy velocities, as well as the range of ram pressures calculated by 3D hydrodynamical simulations.  
  Therefore, our estimates are consistent with the scenario of ram pressure creating the observed deficit regions.  

\item
The internal relativistic plasma pressures range between $\sim$2 and $16\times 10^{-12}$~dyn~cm$^{-2}$ and appear to be lowest in some of the galaxies with the strongest radio deficits.  
Thus the strength of the radio deficit may also be affected by the internal relativistic plasma pressures of the galaxies as well as external ram pressure.
Further work will be required to learn how much the radio deficits depend on these 2 variables.  

\item
  The global radio/FIR ratios of these cluster galaxies are systematically higher than the average value found for field galaxies and appear to increase with increasing severity of the local radio deficit, which may trace current ram pressure.  
   We attribute this to an increase in the CR energy density which is greater when the interaction between the ICM-ISM is more energetic.  
      We also show that the amount of mechanical luminosity available from a galaxy's motion through the ICM is significantly larger than what is required to heat the CR plasma.    

\end{enumerate}

This paper describes a preliminary study comparing the FIR and radio
emission distributions of Virgo cluster galaxies using the first 10
out of $\sim$40 galaxies for which {\it Spitzer} data was available.  
We have clearly shown that by comparing the FIR and radio continuum
distributions of these cluster galaxies we can address a number of
physically important characteristics affecting their evolution
including the strength and directions of the ICM wind. 
The inclusion of the remaining $\sim$30 sample galaxies along with
other ancillary data sets will help to better distinguish among scenarios presented here and better illuminate our understanding of how the ICM affects the relativistic phase of the ISM.

\acknowledgements
We would like to thank the anonymous referee for their useful suggestions that helped to improve the content and presentation of this paper.  
We are grateful to the SINGS team for producing high quality data sets used in this study.  
This work is based in part on observations made with the {\it Spitzer}
Space Telescope, which is operated by the Jet Propulsion Laboratory,
California Institute of Technology under a contract with NASA. 
Support for this work was provided by NASA through an award issued by
JPL/Caltech.

\end{document}